\def \Aa {\AA{ }}

\def \Ia {\AA$^{-1}${ }}
\def \IA {\AA$^{-1}$}
\def \ybc {YBa$_2$Cu$_3$O$_7$}
\def \ybcs {YBa$_2$Cu$_3$O$_7${ }}
\def \CuOs {CuO${_2}${ }}

\def \tl {Tl$_2$Ba$_2$CuO$_{y}$}
\def \tls {Tl$_2$Ba$_2$CuO$_{y}${ }}
\def \Tc {T$_{c}$}
\def \Tcs {T$_{c}${ }}
\documentstyle[aps,epsf]{revtex}
\tightenlines

\begin{document}
\draft
\title{
Correlated local distortions of the TlO layers  in \tl:\\
An x-ray absorption study \\
}
\author{G. G. Li, F. Bridges }
\address{
Physics Department, 
University of California,
Santa Cruz, CA 95064
}
\author{J. B. Boyce}
\address{
Xerox Palo Alto Research Center, Palo Alto, CA 94304
}
\author{T. Claeson}
\address{
Physics Department, 
Chalmers University of Technology, 
S-41296 Gothenburg, Sweden
}
\author{C. Str\"{o}m, S.-G. Eriksson}
\address{
 Department of Inorganic Chemistry,
 Chalmers University of Technology, 
S-41296 Gothenburg, Sweden
}
\author{S. D. Conradson}
\address{
MEE-11, Los Alamos National Laboratory, Los Alamos, NM 87545
}
\date{April 18, 1994}
\maketitle
\begin{abstract}
We have used the XAFS (x-ray-absorption fine structure)
technique to investigate the local structure
about the Cu, Ba, and Tl atoms in orthorhombic \tls with a superconducting
transition temperature \Tc=60 K. Our results clearly show that the
O(1), O(2), Cu, and Ba atoms are at their ideal sites as given by
the diffraction measurements, while the Tl and O(3) atoms are more
disordered than suggested by the average crystal structure.  The 
Tl-Tl distance at 3.5 \Aa between the TlO layers does not change, but the 
Tl-Tl distance
at 3.9 \Aa within the TlO layer is not observed and the Tl-Ba and Ba-Tl 
peaks are very broad.
The shorter Tl-O(3) distance in the TlO layer is about 2.33 \AA, significantly
shorter than the distance calculated with both the Tl and O(3) atoms at
their ideal $4e$ sites ( $x=y=$0 or $\frac{1}{2}$).
 A model based on these
results shows that the Tl atom is displaced along the $<110>$ directions from
its ideal  site by about 0.11 \AA; the displacements of neighboring Tl
atoms are correlated.  The O(3) atom is shifted  from
the $4e$ site by about 0.53 \Aa roughly along the $<100>$ directions.
A comparison of the Tl L$_{III}$-edge XAFS spectra from three samples, 
with \Tc=60 K, 76 K, and 89 K, 
shows that 
 the O environment around the Tl atom is sensitive to \Tcs  while
the Tl local displacement is insensitive to \Tcs and the
structural symmetry. 
  These conclusions are compared with other
experimental results and the implications for charge transfer and
superconductivity are discussed.

\end{abstract}

\section{Introduction}
The Tl$_m$Ba$_2$Ca$_{n-1}$Cu$_n$O$_{2(n+1)+m}$ 
system is a model family of high-temperature superconductors.  The
structures consist of copper perovskite-like blocks containing $n$
\CuOs planes ($n$ up to 6 has been observed) separated by one or two TlO layers. 
  Each CaO layer is
sandwiched between two \CuOs layers while the BaO layer is sandwiched
between the \CuOs and TlO layers.  For Tl-monolayer ($m=$1)
 phases, the superconducting transition temperature (\Tc) increases 
to 122 K  with the number
of \CuOs planes from $n=1$ to $n=4$ and then decreases with $n=$5 to 6;  
for Tl-bilayer phases, \Tcs increases up to 125 K with the number of \CuOs
planes $n$ from 1 to 3 and then decreases with $n$=4, 5.  The fully
stoichiometric Tl-bilayer Tl$_2$Ba$_2$Ca$_{n-1}$Cu$_n$O$_{2n+4}$ compounds
would contain no carriers if the formal ionic valence states,
Tl$^{3+}_2$Ba$^{2+}_2$Ca$^{2+}_{n-1}$Cu$^{2+}_n$O$^{2-}_{2n+4}$, are assigned,
and thus  would not be superconductors. However, there is a hybridization of Tl
$6s$ and O $2p$ orbitals which either touch the Fermi surface or cross
it\cite{Agrawal93,Kasowski}, so the Tl does not act as a completely $3+$ ion and
charge transfer might occur between the \CuOs and TlO layers.  Other effects
can also change the hole-doping in the system including: 
(i) excess or missing O;  
(ii) Ca$^{2+}$ and/or Cu$^{2+}$ substitution for Tl$^{3+}$; (iii) Tl deficiency; 
(iv) ordering of the O and/or Tl atoms  in the TlO layers.
Which doping mechanism primarily controls  \Tcs is still an unsolved problem.
 Structural studies on these compounds
 will be helpful in understanding the mechanism of 
superconductivity in this system.

\tls (Tl-2201)\cite{Sheng} is particularly suitable for this study
 because it has a simple structure (calcium-free and only 1 \CuOs layer per
formula); the highest \Tcs  is about 90 K. 
The structure of Tl-2201, in fact, has been
extensively studied since its first discovery in 1988\cite{Sheng}. Reasonable
agreement for the general structure  has been achieved from
diffraction 
studies\cite{Torardi,Parise88,Hewat88,Parise89,Shimakawa,Kolesnikov,Liu,Manako,Strom93}. 
It contains Tl$_{2}$O$_{2}{ }$ double layers and a
\CuOs plane separated by a BaO layer, as shown in Fig. 1.  
For convenience of description, we define the $4e$ site ($x=y=$ 0 or
$\frac{1}{2}$) as the ideal site for Tl and O(3) in tetragonal 
\tls and treat
orthorhombic \tls to be a pseudo-tetragonal structure.
The atomic positions of Tl and O in the Tl$_{2}$O$_{2}{ }$ double 
layers, however, are not well defined,
due to a large anisotropic thermal factor and/or various static displacements
from their ideal sites.  Most of the refined structures suggest a
distorted  site for the O(3) with a displacement of $\sim$0.31-0.42 \Aa along the
$<100>$ direction\cite{Torardi,Shimakawa,Liu,Manako} or  $\sim$0.22-0.28
\Aa along the $<110>$ direction\cite{Parise88,Parise89,Kolesnikov}, 
while some of them show that the Tl atom is also displaced from its ideal  
site by $\sim$0.11-0.14 \Aa along the $<110>$ direction\cite{Parise89,Kolesnikov} 
or $\sim$0.16 \Aa along the $<100>$ direction\cite{Liu}.  The 
refined thermal factors in these studies are usually quite large in 
the ab-plane for the Tl and O(3) atoms.  
Recently, it has been made clear that the thallium content has a strong
influence on the lattice symmetry and thus on the average atomic position in the
TlO layer\cite{Strom93,Shimakawa93,Strom}
Furthermore, significant distortions have recently
been observed in both \tls and Tl$_2$Ba$_2$CaCu$_2$O$_8$ (Tl-2212) by means of 
 x-ray-absorption fine structure (XAFS)\cite{LiTl}. Similar
distortions  are found in
Tl-2212 by the pair-distribution-function (PDF) analysis of neutron 
scattering data\cite{Dmowksi,Toby89}.  
It appears that these distortions not only have short-range correlations, but 
 may in some cases
also have long-range correlations, as evidenced by the presence of
the modulational structure for selected sections, observed in
 high-resolution electron microscopy and electron
diffraction\cite{Parkin,Fitz,Oku,Beyers,Nakajima93}.  In addition, the 
temperature dependence of the local structure in the \CuOs plane has 
been studied by XAFS in
Tl-2201\cite{LiO,Yamaguchi} and by PDF in Tl-2212\cite{Toby}, respectively.
The splitting of the Cu-O(axial) in TlBa$_2$Ca$_3$Cu$_4$O$_{11}$ has also
been investigated by XAFS in the range 10$<$T$<$156 K\cite{Allen91}.

In this study, we have focused on the static distortion of the \tl, 
especially in
the TlO layer.  We have used the XAFS technique\cite{Hayes,Stern} to probe the
local structure around {\it all} the metal atoms, Cu, Ba, and Tl in \tl, to 
revise the earlier results based on the Tl L$_{III}$ edge studies\cite{LiTl}.  
Both the sample preparation and the XAFS data analysis are
slightly different from the previous study\cite{LiTl};
they are described in Sec. II. 
The XAFS data analysis  (presented in Sec. III) uses
constrained fits\cite{LiCo} to reduce the number of independent 
parameters and to check the previous model.  
The XAFS data and
the data analysis for the Cu and  Ba K-edges, and the Tl L$_{III}$-edge are
 discussed
in Secs. IV and V, respectively.  A discussion and a comparison of our
results with other experimental observations are given in Sec. VI, and the results
summarized in Sec. VII.

\section{Experimental Details}

\subsection{Sample Preparation and Characterization}
Samples of about 10 grams weight were prepared from a mixture of Tl$_2$O$_3$,
Ba(OH)$_2\cdot $H$_2$O and Cu(NO$_3$)$_2\cdot $nH$_2$O, always in the ratio of
the cations desired in the final material with no excess of any element.  The
powders, milled in ethanol and thereafter dried in an inert atmosphere (N$_2$),
were heat treated by a two-step sintering process.  The heating rate in both
steps was 180$^o$ C/hour and the temperature was regulated by a controller,
Eurotherm 818P, with a temperature stability better than $\pm$0.5$^o$ C.

In sintering step I the mixtures were heated in Al$_2$O$_3$ crucibles with
Al$_2$O$_3$ covers at 760$^o$ C for 3 hours with only a limited loss, if any at
all, of thallium.

In sintering step II pellets were placed in Al$_2$O$_3$ crucibles sealed with
gold gaskets and cover.  They were subsequently centered in small pieces
of ceramic tube and fixed with steel wedges (a procedure that prevents
evaporation from the crucible) and heated at 860$^o$ C for 6 hours.  The loss
of weight in sintering step II, if attributed solely to thallium evaporation,
is below 0.3\% of the total amount of thallium.  There is no sign of substitution
of gold into the material or of incorporation of thallium into the gold
gasket.  From EDX (energy dispersive x-ray spectroscopy) analysis,  no traces of
Al or Au can be detected in the sample.  For estimated errors in barium and
copper content see Ref. 12.

Sample purity and lattice parameters were determined from Guinier film data
using silicon (NBS 640b) as an internal standard.  A computerized
photoscanning system in combination with programs for indexing and unit cell
refinements were used for the extraction of lattice parameters (a=5.4881(2)
\AA, b=5.4532 \AA, c=23.216(1) \AA).  The impurity content was below 2\%.

Neutron powder diffraction data were collected at the pulsed neutron source
ISIS, RAL, UK using the high flux, medium resolution powder diffractometer
POLARIS.  A least-squares profile refinement program, TF12LS, based on the
Rietveld technique was used to analyze the TOF (time-of-flight) neutron powder
diffraction data.  The refinement was carried out in space group Abma and 
selected bond distances are found in Tables I-III.

The superconducting properties were studied by ac susceptibility measurements
using a Quantum design SQUID magnetometer  at temperatures down
to 4 K in a field of 50 Gauss.  \Tcs was determined to be 60 K.  For further
details concerning sample preparation and characterization consult Ref. 12.

\subsection{X-ray absorption measurements}
The \mbox{XAFS} samples were prepared by brushing a fine powder
($\approx $ 30 $\mu m$) onto Scotch tape.  
Each layer  was selected to be free of pinholes to minimize the XAFS
glitches\cite{Glitch} and several layers were stacked to obtain a
sample with a thickness of approximately one absorption length.  Samples
were measured at 80 K using a liquid-nitrogen dewar.  Several data sets were 
collected for each sample to
check for consistency and averaged to improve the signal-to-noise ratio.   
The x-ray absorption data were collected on  beamline 10-2 at
the Stanford Synchrotron Radiation Laboratory using Si (400)
monochromator crystals.  The double monochromator was detuned on its
rocking curve to approximately one half of maximum transmitted x-ray
intensity to reduce the harmonic content of the beam.

\section{XAFS data analysis technique}
\subsection{Standard Procedures}

The standard procedures for the reduction and analysis of the \mbox{XAFS} data 
are described elsewhere \cite{Hayes,LiCo}.  The extracted XAFS function,
$k\chi (k)$, is given by
\begin{equation}
k\chi(k)=\sum_{j}\frac{N_{j}F_{j}(k)}{R_{j}^2} sin[2kR_{j}+\phi_{j}(k)]exp(-2k^{2}
   \sigma^{2}_{j}-2R_{j}/\lambda)
\end{equation}
where the sum is taken over shells with $N_{j}$ atoms at a distance 
$R_{j}$ from the absorbing atom, $k$ is the momentum of the photoelectron
given by $k=[2m(E-E_{o})]^{1/2}/\hbar$ where  $E_{o}$
is the energy at one-half the absorption edge height, 
$F_{j}$(k) is the back-scattering amplitude,
 $\phi_{j}(k)$ is a phase shift of the photoelectron due to its interaction
 with the
back scattering and absorbing atoms, $\lambda$ is the effective electron
mean free path, and
$\sigma_{j}$ (Debye-Waller factor) is the average variation (both static
and dynamic) of $R_{j}$. Examples of the
k-space data are given in Fig. 2 for the Cu and Ba K-edges and Tl 
L$_{III}$-edge.   A Fourier transform of these k-space data exhibits
peaks in r-space which correspond to neighbors at different radial
distances from the absorbing atom.

 To obtain numerical values for $R_{j}$, $N_{j}$ and
$\sigma_{j}$, we perform iterated least-square fits of the real and
imaginary parts of the Fourier transform of $k\chi(k)$ to a sum of standards 
over a particular 
 range in real (r) space. Identical k-space transforms are used for 
the sample and the standards. In these fits, the
overlaps of the radial distributions of other neighbors in r-space are included.

It is difficult to extract all the isolated, single pair  standards needed
for this analysis from other experimental standards.  In this study  all the 
standards are generated by the FEFF5 code\cite{FEFF}, using a cluster as 
close to the real lattice as possible with a radius of about 7.5 \Aa
from the central absorbing atom to the boundary scattering atoms. The
multiple scattering paths are included in the fit when it is necessary.
In earlier studies we have used experimental atom-pair standards
extracted from data collected for standard compounds.  We have compared
our experimental standards with FEFF5 and have obtained very good agreement 
for about twenty pairs\cite{LiFeff}.  Since a complete set of experimental
standards was not available, we decided, based on our recent study,
to use FEFF5 standards.
One of the  problems in using the standards from FEFF5 is the uncertainty
of the amplitude-reduction-factor, $S_o^2$.  We use two different values
of $S_o^2$ which we have obtained from our detailed comparisons\cite{LiFeff};
 0.8 when the back-scattering atom is O, and 0.9 when the 
back-scattering atom is a metal atom.  

\subsection{Background-feature removal}
The standard background removal procedure which uses a simple spline or
polynomial for $\mu _o(E)$,  does not work well when the background consists
of some features due to multielectron excitations\cite{Li92,Booth94} and
Ramsauer-Townsend-like effects\cite{Booth94,Rehr94}.  These features have
been found at the Br (in RbBr), Kr, Rb (in RbBr), Ba (in BaO), 
Ce (in CeO$_2$), 
and Pr (in PrBa$_2$Cu$_3$O$_7$) K-edges\cite{Li92,Booth94,Rehr94}
and the Au, Tl (in Tl$_2$O$_3$), Pb (in $\beta -$PbO$_2$), and Bi (BaBiO$_3$)
 L$_{III}$-edges\cite{LiFeff,Li92} in our studies.  Similar features
have also been observed and confirmed by other 
groups\cite{Frenkel93,Filipponi93,DAngelo93}.  An iterative background
extraction and removal procedure has been proposed to obtain and
 remove these features in the previous work\cite{Li92}.

When the simple standard background removal procedure was applied to the
Ba K-edge and Tl L$_{III}$-edge XAFS data of \tl, spurious humps appear in
the low-r region of the Fourier transformed data as shown in Fig. 3.
Instead of performing the iterative background removal procedure  to
these complicated data directly, we utilized the extracted background from the
BaO\cite{Rehr94} and Tl$_2$O$_3$\cite{LiFeff} data; 
the new background was obtained by fitting the
corresponding extracted background plus simple spline functions to the
data.  These new backgrounds are plotted in Fig. 3.  They contain more
wiggles than the background obtain from the standard procedure but
result in smoother r-space data in the low-r region
 than those from the standard procedure.  This simple method works very
well.  We note that using more spline sections in the standard procedure
may also  minimize the humps
in the low-r region but that usually  reduces the amplitude of 
the first r-space peak,  simultaneously.

\subsection{Constrained Fits}
One of the common problems in the XAFS data analysis is that there are too many
fitting parameters if the fit includes further shells in a complicated
system. For example, if the fitting range is 3.0 $\sim$ 13.0 \Ia in
k-space and 1.0 $\sim$ 4.0 \Aa in r-space, the {\it estimated}
allowed number of  
fitting parameters is $\sim$21, according to the formula 
$2 \Delta k \Delta r/\pi +2$\cite{Stern93}.
In a complicated system, such as the Cu edge data in 
\ybc\cite{LiCo,Bridges89,Boyce89}, the number of atomic pairs could
exceed 8, which needs 24 or more fitting parameters, even if all 
the E$_o$' are fixed and
not considered as fitting parameters.  Therefore, a constrained fit is
required in the XAFS data analysis for a complicated system. These
constraints include using the lattice constants 
and the known average locations of the atoms in the unit cell from
diffraction measurements,  combining  the XAFS results viewed by different
atoms, and using equations which relate a parameter to one or more other
parameters (for examples, keeping the ratio
of some parameters constant; setting a parameter to the weighted sum of
other parameters) to check possible models. The 
constraints force the fit towards a particular minimum in
parameter space, and the fit is not easily trapped in some other local
minima.  In this way, the constrained fits can help reject  results
which are physically inconsistent and to check particular models for the local
structure.  This method has been used to a limited extent in our earlier
work\cite{Bridges89,Boyce89} and more extensively in our recent
studies\cite{LiCo}. 

\section{XAFS Data Presentation and Qualitative Features}

\subsection{Cu environment}

From diffraction 
studies\cite{Torardi,Parise88,Hewat88,Parise89,Shimakawa,Kolesnikov,Liu,Manako,Strom93}, the average local environment about a Cu atom is
as follows: 4 O(1) neighbors at r=1.93 \Aa in the \CuOs plane, 2 O(2)
neighbors at 2.71 \Aa along the c-axis, 8 Ba neighbors at 3.35 \Aa above and
below the \CuOs plane, and 4 Cu neighbors at 3.87 \Aa in the \CuOs plane. 
In XAFS data, the
Fourier transform of the XAFS function $k\chi (k)$ to real (r) space
is a sum of peaks corresponding to each of these different neighbors;
however, each distance is the average local value, not necessarily the 
value calculated from the average unit cell.  The question considered in
this section is how well does the Cu local environment agree with
diffraction results.   

The Cu K-edge XAFS data in r-space
are shown in Fig. 4.  The transform range is a square window from 3.1 to
16.5 \IA, Gaussian broadened by 0.3 \IA.  Both the real part and the
magnitude of the complex transform are plotted.  
The distances between the absorbing atom and the back-scattering atoms
are determined by the phase of the XAFS function,
which is well represented by the positions of the zero-crossing points
of the real and imaginary parts of the Fourier transform. 
However, there is a shift,
$-0.2\sim -0.5$ \AA, of the XAFS peaks from the real positions due the
photoelectron phase-shifts. 

To compare the local structure which determines the  Cu K-edge XAFS
with the average structure obtained from diffraction, we generated the
theoretical XAFS function using FEFF5\cite{FEFF} and the refined crystal
structure obtained from diffraction measurements\cite{Torardi}.  
In the calculation,
all the atoms within a distance of 7.5 \Aa from the 
central absorbing atom,
 and all the scattering paths with a contribution of 4\% or more,
relative to the maximum contribution, are included. 
To roughly simulate both the static and thermal disorder in the calculation,
we have set the sample temperature to 100 K and the Debye temperature 
to 700(K \AA)/(0.5R$_j$) which increases $\sigma$ with distance R$_j$ 
 for more distant shells.  (The same thermal 
parameters are used for the other edges in this paper unless noted).  The
Fourier transform of the theoretical XAFS (without any fitting) is shown 
in Fig. 4 by dotted lines,
with the same k-space  window as for the experimental data,
 an overall amplitude-reduction-factor of 0.9, and an overall E$_o$ shift
 of $-2.0$ eV.  

Fig. 4 shows that the phase
of the theoretical XAFS agrees very well with that of the
experimental data up to 6.5 \AA.  This means that the O(1), O(2),
Ba, and Cu atoms  are well ordered about a given Cu atom, in excellent
agreement with the diffraction results.
We point out that the Cu edge XAFS data is not sensitive to the distortion
of the distant TlO layers if there is any, especially when the distortion is
in the ab-plane, because the TlO and \CuOs layers are separated by 4.7 \Aa
and a small splitting of the Cu-Tl and/or O(3) distances can be absorbed by a
large Debye-Waller factor.  We note that the subtle differences in
zero-crossing points are due mainly to the errors in E$_o$, especially in the 
region below 2.3 \Aa where the O atoms make a major contribution.
The nearest O neighbors directly touch and partially screen  
the excited absorbing atom 
which has a hole in the  core-shell level, thus the E$_o$ shift for
the nearest O neighbors should be different from the distant neighbors.  
Since the same E$_o$ shift is used for all neighbors in the simulation, 
an error in the phase-shift  can be introduced in the first O peak. 
 The discrepancy in the magnitude is mainly from the
rough estimation of the Debye-Waller factors and the
amplitude-reduction-factors of the XAFS function in the theoretical
calculation and is not strongly related to the separations of the atomic pairs.

We emphasize that in the theoretical calculation we did not adjust the XAFS 
parameters, N$_j$, R$_j$, $\sigma_j$, and E$_{oj}$, individually; instead,
we only adjusted the overall amplitude and E$_o$  and  simulated the
Debye temperature of the different pairs by a simple formula.  This method 
makes it  possible and easy
to compare the experimental XAFS data with a theoretical function calculated
for a particular model up to 7 \Aa (for example, from  the refined 
average unit cell obtained from diffraction).   However,  it is still difficult 
to  make a detailed fit to the experimental XAFS data much beyond 4 \AA.
 
\subsection{Ba environment}

The Ba atom is in a plane between the \CuOs and TlO layers.   From diffraction 
measurements\cite{Torardi,Parise88,Hewat88,Parise89,Shimakawa,Kolesnikov,Liu,Manako,Strom93}, 
it has 4 O(1) neighbors
at 2.73 \AA, 4 O(2) neighbors at 2.84 \AA, 1 O(3) neighbor at 3.00 \AA,
4 Cu neighbors at 3.35 \AA, 5 Ba neighbors (4 in the ab-plane and 1
along the c-axis)  near 3.86 \AA, and 4 Tl
around 3.9 \AA\cite{Torardi}.  Again we ask the question if the local
structure about the Ba atoms is the same as that given by
the diffraction results assuming no distortions of the Tl atoms.
The Fourier transform of $k\chi(k)$ for the Ba K-edge data is shown in 
Fig. 5; window is from 3.5 \Ia to 15.5 \Ia, Gaussian broadened by 0.3 \IA.  
We have generated the theoretical Ba XAFS function following the procedure
described earlier for the Cu edge; it is plotted as dotted lines in Fig. 5a.  

The phase of the theoretical and experimental XAFS function
agreed very well up to 3.7 \Aa (a small discrepancy in the phase of the first
peak can be easily minimized by using a different E$_o$ shift). This again 
indicates that the 
O(1), O(2), Cu and Ba atoms are well ordered.  Similarly, there is 
excellent agreement in the phase for a large peak
  around 5.2 \Aa between the simulation and experimental XAFS
data.  The dominant contributions to this peak arise from the collinear,
forward, multiple scattering 
of the Ba-O(1)-Ba and the multiple scattering of intraplanar Ba-O(2)-Ba pairs.
The phase of the real part of the transform within the 5.2 \Aa peak is
shifted by nearly $\pi /2$ relative to the phase that would be observed for 
a single scattering Ba-Ba peak ({\it e.g.} as occurs near 3.5 \Aa in Fig. 5)
 due to the forward multiple scattering contributions.
The good agreement between the calculations and the experimental data 
for this peak strongly suggests that the Ba atom is well ordered.  

More importantly, however, there is a significant difference
 between the calculations and the experimental data on the high-r side
of the peak near 3.6 \AA (see Fig. 5c in which a small fraction of the data
is shown  with an enlarged $x-$axis), where the 
Ba-Tl pair (r= 3.90 \AA) should make a major 
contribution.  The experimental data in this region of r-space
can be simulated much better by 
FEFF5 when  the Ba-Tl pairs are excluded; {\it i.e.,} the net Ba-Tl peak
is small.    This
means that the Ba-Tl peak is either very broad due to disorder of
the Tl atoms or is split into multi-peaks due to a displacement of the Tl
atom from its ideal site. 

\subsection{Tl environment}

There are two adjacent TlO layers, which are surrounded by two BaO layers, one
above and one below. The average local structure around the Tl atoms obtained
from diffraction\cite{Torardi,Parise88,Hewat88,Parise89,Shimakawa,Kolesnikov,Liu,Manako,Strom93} consists of 1 O(2) and 1 O(3) atoms near 2.0 \AA, 4 O(3)
within 2.3-3.2 \AA; 4 interplanar Tl atoms near 3.5 \Aa    
and 4 intraplanar Tl atoms near 3.9 \AA; and 4 Ba atoms near 3.9 \AA.
However as outlined in the introduction and the above discussion  at the Ba
edge, there is considerable evidence that the TlO layers are distorted. 
Here we investigate this disorder using Tl edge XAFS data. 

First,  we  performed a 
multiple-scattering-cluster-simulation for the Tl L$_{III}$-edge XAFS
(as described above for the Cu K-edge data)
assuming the structure of Ref. 4 in which the Tl is at the ideal $4e$ site
while the O(3) is displaced along the $<100>$ direction by 0.37 \Aa from its
ideal $4e$ site. Both the simulated and experimental XAFS,
$k\chi(k)$,  were Fourier transformed to r-space using a k-space
 range of 3.2 - 15.5 \IA, 
Gaussian broadened by 0.3 \IA,  as shown in Fig. 6a.
In the simulation, we used two different sets of Debye-Waller temperatures,
300 (K \AA)/(0.5$R_j$) for Tl-Tl pairs and 700 (K \AA)/(0.5$R_j$) for other
pairs, to better simulate both the static and thermal disorder.  This
gives a good simulation above 4.5 \Aa in r-space. 

Significant differences in the phase between the simulated and 
experimental data are observed in Fig. 6a.  Thus at least some of 
the distances  derived from the average, crystallographic structure are not
consistent with these XAFS data.  For the first O neighbor
peak (1.3--2.3 \Aa in Fig. 6) of the simulated data, the 
phase of the low-r side (1.3--1.8 \AA),
where the Tl-O(2) and Tl-O(3) pairs at r=2.0 \Aa make their major contributions,
agrees reasonably well with the  experimental data; however, at the high-r side 
of this peak (1.8--2.3 \AA) , where the
Tl-O(3) pair at r=2.5 \Aa makes its major contribution, the phase is not
consistent with the experimental data and a contribution in the simulated
data is missing in the experimental data.  A comparable inconsistency is
observed in the peak centered near 3.6 \AA.  
 The phase and the shape of the simulation for this peak 
agree well with the actual data in the low-r side, where the interplanar Tl-Tl
pair makes its major contributions; however, it does not agree in the
high-r side (3.7--4.0 \AA), where the intraplanar Tl-Tl and also the
 Tl-Ba pairs make their
major contributions.  Furthermore, the amplitude of the simulated peak in
r-space near 3.6 \Aa is almost twice as big as that in the data.  This
discrepancy is too large to be explained just by the errors in the
amplitude-reduction-factor and the Debye-Waller factors. This implies a
significant distortion and/or disorder in the Tl-Tl and Tl-Ba distances from
the average distances obtained from the refined structure in diffraction.
A simulation without the intraplanar Tl-Tl pairs
bears a much closer resemblance to the experimental spectra near 3.5 \AA. 
These comparisons 
again suggest that  the Tl atoms, and probably
also the O(3) atoms,  occupy more distorted sites 
in the ab-plane  than those obtained from the average 
diffraction experiments.  

\section{Detailed XAFS Data Analysis}

From the qualitative analyses of the Cu, Ba, and Tl edge XAFS data
shown above, we conclude that the positions of the Cu, O(1), O(2), and Ba atoms
in the unit cell are consistent with the average crystallographic
structure obtained from diffraction studies.  However, the Tl and
O(3) atoms are significantly more distorted than the average structure
obtained in many diffraction studies.  The XAFS data indicate that these
distortions are  located in the ab-plane of the TlO layer as 
suggested in some diffraction investigations.  
These qualitative features are model independent and give a sound
guidance for further structural analysis.

In this section, we will perform  nonlinear-least-square fits to the XAFS
data to  further check the qualitative features shown above and gain
quantitative information on the local structure about the Cu, Ba, and Tl
atoms.

\subsection{Cu environment}

The r-space Cu K-edge XAFS data have been fit up to 3.9 \Aa using individual
standard atomic pairs generated by FEFF5. Five pairs are included in the
fit: Cu-O(1) near 1.9 \AA, Cu-O(2) near 2.7 \AA, Cu-O(1)
near 4.3 \AA, Cu-Ba near 3.4 \AA, and Cu-Cu near 3.9 \Aa in which the
 multiple-scattering-paths for Cu-O(1)-Cu are included. The experimental
data and the fit are shown in Fig. 7, where the same Fourier transform
range is used as in Fig. 4 and the fitting range is between 1.0-3.9 \AA.
Overall, a high quality fit has been achieved.  A double-well potential
was used in Ref. 23 to improve the fit for the first Cu-O(1) peak; 
  however, in this study, the refinement of the first peak is not important 
and we use only a single Cu-O(1) peak. 
The structural parameters extracted from the 
XAFS curve-fitting are tabulated in Table I, together with
the results from diffraction refinement. Our XAFS results are in excellent
agreement with the diffraction results, as expected from our earlier
comparison of the simulated and experimental XAFS r-space spectra, in which
there was an excellent phase agreement up to 6.7 \AA.  These XAFS results
are also very close to the previous results\cite{LiSr}.
 
\subsection{Ba environment}
In Sec. IV,  we showed that the Tl is disordered or distorted.
So in the fits to the Ba K-edge data, we considered three situations for
comparison: (1) no Ba-Tl peak, (2) a single Ba-Tl peak, and (3) a split
Ba-Tl peak. For the first case, we
fit the Ba K-edge XAFS data with 3 O peaks (at r=2.7 \AA, 2.8 \Aa
and 3.0 \AA, respectively), 1 Cu and 1 Ba peak.
The Tl peak near $r$=3.9 \Aa was not included.  The Debye-Waller
factor of the Ba-Cu peak was fixed to have the same value as that of the
Cu-Ba peak from the Cu K-edge fit.  A reasonably good fit is achieved
and all the extracted parameters are reasonable, which means that the Ba-Tl 
pairs do not make a significant contribution to the Ba edge XAFS, in
contradiction to what one would expect for the undistorted Tl layers.

Secondly, we included the Ba-Tl peak in the fit and fixed the amplitude
 of that peak to 4, the number predicted by diffraction.
The quality of the fit was improved by about 25\% and the extracted
distances and the amplitudes for the Ba-O, Ba-Cu and Ba-Ba peaks remained
essentially unchanged. The Ba-Tl distance is about 3.90 \AA, very much
the same as
the average distance given by diffraction. The Debye-Waller factor of
the Ba-Tl peak is about 0.094 \AA, much bigger than that of the Ba-Ba peak
(0.066 \AA).

Finally, we tested the model proposed earlier\cite{LiTl}.  Here we require
consistency between the XAFS
spectra collected from three different absorbing  atoms, achieved by using
constraints in the fitting procedure.
Following our previous discussion, we assumed that the Tl atoms are shifted
along $<110>$ so that the 4 Ba-Tl pairs split into three distances, with
roughly 1 pair at $r-\delta r$, 2 pairs at $r$ and 1 pair at $r+\delta r$.   
Therefore, we included 3 O, 1 Cu, 1 Ba and three Ba-Tl peaks in the fit.
To minimize the number of fitting parameters and to check the model,
we fixed the average distance $r$ and the amplitudes of these three
peaks to the values predicted by the model and only allow $\delta r$ to
vary for  these Ba-Tl peaks.  The Debye-Waller factors of the three Ba-Tl
peaks and the E$_o$ shifts for the same type of  back-scattering atoms
 were also kept equal.  
In this way, the number of the variables was reduced to 18, well below
23, the estimated number of allowed variables obtained from
($2\Delta r \Delta k /\pi +2$)\cite{Stern93}.
The quality of the fit and the extracted
parameters for O, Cu and Ba peaks are essentially the same as the fit
with a broader Ba-Tl peak but here $\delta$$r$=0.10 \AA.  As a final step, 
we also allowed the average
Ba-Tl distance and the amplitudes of the Ba-Ba and Ba-Tl peaks to vary.
The final fitting parameters remained essentially the same as the ones with
the strong constraints.

The experimental spectra and the final fitting
curves are plotted in Fig. 8a; the contributions from O, Cu and Ba
atoms are shown in Fig. 8b, while the contributions from the split Ba-Tl
pairs are shown in Fig. 8c.  These results  clearly show that the 
contributions of the Ba-Tl pairs are quite small compared to the Ba-Cu and
Ba-Ba pairs.  The detailed results 
 are listed in Table II,
together with the results from diffraction and the model for
comparison.  We would like to point out that the Ba-O(3) distance is not
sensitive to the displacement of the O(3) atom in the ab-plane. 
For Ba-Tl peaks, $\delta r$ is slightly coupled with the Debye-Waller
factor $\sigma$ and the coordination number $N$.  It varies from 0.08
\Aa to 0.11 \Aa depending on $\sigma$ and $N$ (or $S_o^2$).
We also used different k weighting ({\it i.e.}, $k^n \chi (k)$ with n=1, 2, 3)
and different k ranges for the Fourier transform in the fit.  The derived
parameters are essentially the same within the estimated uncertainty.   
In all cases, the splitting distance, $\delta r$, for the
Ba-Tl pair is significantly shorter than 0.2 \AA, the distance predicted 
by the previous model. This
means the displacement of   the Tl atom in the previous model (0.28 \AA) is too
large and has to be modified to fit the Ba edge XAFS data.

We also simulated the Ba K-edge XAFS data with a distorted model for the
TlO layer.  Here we modified the previous model by assuming that the Tl atom is
displaced 0.11 \Aa from its ideal site along the $<110>$ direction.
 The simulated data are plotted in Fig. 5b.   The agreement with the data
is much  improved for the r-space range 3.5--4.5 \Aa  
compared to the earlier simulation (Fig. 5a), as
illustrated  in Fig. 5c.    There is also improved agreement in phase 
near 6 \AA.

\subsection{Tl environment}
From the analysis of the Cu and Ba K-edge XAFS spectra, we conclude that
the O(1), O(2), Cu, and Ba atoms are at their ideal sites in
the unit cell just as diffraction shows, but the Tl atom is
shifted from its ideal  site.  However, it is difficult,
 from the Cu and Ba edge XAFS spectra only,  to distinguish between a randomly
 disordered Tl site (which gives a broad peak) and a distorted Tl site
(which gives multiple peaks) 
and hence to derive the actual displacement of the Tl atom if it is at a
distorted site.   Further information from the Tl edge is desirable.

The Tl L$_{III}$-edge XAFS data are more sensitive to the Tl distortion, but 
the data analysis is more complicated by the fact that there are two adjacent TlO
layers.  In the fits to the Tl edge data, we considered 
the Tl displacements along the $<110>$ direction with two different magnitudes:
(1) $\sim$0.3 \AA, similar to the
previous model\cite{LiTl}, and (2) $\sim$0.1 \AA, as suggested by the Ba
edge XAFS data.  In comparisons with diffraction studies, we also tried
other models,  see Sec. VI.A.  

In order to enable a meaningful analysis of the Tl L$_{III}$-edge data
we treat the Tl atoms as if they all see the same local distortions.  
However, we know that in the related Bi-based cuprate superconductors, 
there exist strong periodic modulations\cite{Tarascon91}
 which result in superstructure with long range
order, a buckling in the ab-plane, and periodically extra O atoms.
Because diffraction results show little disorder for Tl along the
c-axis,  we are motivated to use a simple model for the local
environment.  As discussed later, it is  still compatible with a
superstructure unit cell.

We first performed a rough fit to the data in the region of 
1.3$\sim$3.8 \AA, with 2 O, 1 Tl and 1 Ba peaks, similar to the model 
used in the previous paper\cite{LiTl}. 
The extracted structural parameters are
the following: 1.8 O atoms at 2.04 \AA, 2.9 O atoms at 2.35 \AA, 2.8 (which is too small compared to 6)\cite{Note}
 Tl atoms at 3.51 \AA, and 7.6 Ba atoms 
(but very broad, with $\sigma=0.13 $\AA) at 3.89 \AA.
This shows that the interplanar Tl-Tl pair (at r=3.5 \AA) remains but the
intraplanar Tl-Tl pairs (with an average distance of 3.9 \AA)
 do not make a large contribution to the Tl edge XAFS spectra as would
 occur for an undistorted site.   This result suggests that the Tl atom
is distorted (not randomly disordered) in a particular way so that the 
interplanar Tl-Tl peak remains but the intraplanar Tl-Tl peak seems to disappear.
 The fitting result we obtain here is quite close to the
earlier result\cite{LiTl}.  However, the model based on that result requires
a large shift for the Tl atom which does not fit the Ba edge data as 
discussed above, and the Tl-Tl amplitude is too low.  Therefore,  the previous 
model must be modified to fit both the Ba and Tl edge XAFS data.

Secondly, we followed the previous model but with a smaller shift for 
the Tl atom.
Assuming that the Tl atoms are shifted along the $<110>$ direction as shown in
Fig. 9, the interplanar Tl-Tl distance remains approximately unchanged
 while the intraplanar
Tl-Tl distance splits into two distances.  In addition, the Tl-Ba distance splits
into three distances,  with the average distance unchanged to  first
order.  In order to check this model more carefully, we explicitly
included all the split peaks and the long Tl-O peak at 4.3 \Aa in the 
fit using the constrained-parameter 
technique.  The numbers of neighbors for all
the Tl and Ba peaks are fixed to the values predicted by the model.
The average Tl-Ba and intraplanar Tl-Tl distances are also fixed to the
average distances given by the diffraction measurement, and only 
the split distances are allowed to vary. The Debye-Waller factors of the
three Tl-Ba peaks are set  to be equal to those of the Ba-Tl peaks at 
the Ba edge, 
to further reduce the number of independent variables and to check the
consistency of the fitting results. The E$_o$ shifts for the same
type of back-scattering atoms are kept equal except for the nearest
O atoms.  Thus the total number of variables is 16, 
far below the estimated number of allowed variables, 27\cite{Stern93}.
The fit was carried out in the region of 1.2 \AA$<r<$4.2 \AA, with the
Fourier transform range of 3.3 \Ia $\sim$ 15.5 \IA, Gaussian broadened
by 0.3 \IA.  A high quality fit was achieved.

Finally, we allowed the average Tl-Ba and intraplanar Tl-Tl distances, 
the amplitudes, and the
Debye-Waller for the three Tl-Ba pairs to vary.
 The fitting results remained essentially
the same as those with the constraints.  We also fit the data with different
k weighting ($k^n\chi(k)$, n=1, 2, 3) and k range in the Fourier transform. 
Consistent results were achieved within the estimated uncertainty.

The final  fit is  shown in Fig. 10a.
The contributions from O, Tl and Ba atoms are plotted in Fig. 10, b-d,
respectively.  The contributions from the three individual Tl-Tl peaks
are  plotted below the sum of their contributions. The contribution of
the Tl-Tl at r=4.02 \Aa is small, and  the Tl-Tl peak at
r=3.72 \Aa is  out of phase with the Tl-Tl peak at r=3.50 \AA, resulting in
partial destructive interference.
Thus the sum of the Tl-Tl peaks has almost  the same phase as that of the
Tl-Tl peak at r=3.50 \Aa but  a smaller amplitude, which  
explains why the fit with only one Tl-Tl peak gives quite good
agreement,  but with a
small amplitude.  The three Tl-Ba peaks are also plotted below the sum of
their peaks.  The sum of the three split Tl-Ba peaks is a broad peak,
with a similar phase to that of the Tl-Ba peak at the average
diffraction distance, 3.90 \AA, which again is consistent with one broad
Tl-Ba peak in the previous fit. 

The extracted structural parameters are listed in Table III,
together with the corresponding parameters calculated from the model.
We emphasize that the Tl-Ba distances obtained in this new fit of the Tl 
L$_{III}$-edge data are in good agreement with the Ba-Tl
distances extracted from the Ba edge XAFS data within the estimated
uncertainty.
In the model, we only shifted the Tl atom along the $<110>$ direction to match
the Tl-Tl, Tl-Ba and Ba-Tl distances extracted from our XAFS data. The
best overall agreement is achieved when the Tl atom is displaced by
about 0.11 \AA.  We have  also shifted the O(3) atom along the $<100>$
direction from the $4e$ site by 0.53 \Aa in the model to match
the observed average interlayer Tl-O distance at 2.33 \AA.  The splitting
of the two shorter interlayer Tl-O distances ($r$=2.29 and 2.38) predicted
by the model is consistent with the broad Tl-O peak ($\sigma$=0.131 \AA)
at $r$=2.33 \Aa in the data.
We  point out that
absolute position for the O(3) is not well defined in the data 
because the Tl-O peak at 2.33 \Aa is quite broad  and the Tl atom itself 
is displaced. 
The long interlayer Tl-O(3) peak, predicted to be near $r=$3.12--3.24 \Aa by the
model,  is not clearly observed in the XAFS data, probably because that peak is
considerably broader than the shorter one,
and/or smeared out due to  multiple Tl-O peaks in that range.  Instead of
the well defined pair distances predicted by the model, 
the actual PDF of the Tl-O(3) pairs within the TlO layer may be spread over
the entire range 2.3--3.2 \Aa (refer to Table III) due to the variations of the Tl and O(3)
displacements in both magnitudes and directions.

We again simulated the Tl edge XAFS,  using the FEFF5 code and the 
distorted local structural model (see Fig. 9).  The simulation is shown in Fig. 6b.
The sample temperature and Debye-Waller temperature used here are the same
as those used in the simulation shown in Fig. 6a, except the parameters for
the first two Tl-O pairs ($r$=2.00 \Aa and 2.09 \Aa) in which the
Debye-Waller factors were set to 0.03 \Aa because the static disorder had
already partially  been taken into account by the two separated peaks. 
The new simulation in Fig. 6b shows a great improvement over the old one
shown in Fig. 6a,  and the phase of the new simulation agrees very well with the 
actual data over the whole region.

\section{Discussion}

\subsection{Comparisons with diffraction results}

The extracted distances from our XAFS data analyses
 for the undistorted atomic pairs, such as
 Cu-O, Cu-Ba (Ba-Cu), Cu-Cu, Ba-O(1,2), Ba-Ba, are in excellent 
agreement with those
obtained from diffraction studies, with a typical error of 0.01 \Aa (see
Table I and II).  This
shows the high quality of the XAFS data and the data analyses.  The
extracted distances for the distorted atomic pairs, such as 
Ba-O(3), Ba-Tl (Tl-Ba), Tl-O, are within the range of the distances given
by diffraction investigations (see Table II and III). 

Diffraction studies\cite{Torardi,Parise89,Shimakawa,Kolesnikov,Manako}
 show that the refined thermal factors for the Tl and O(3)  
atoms along the c-axis are only one third or less of those in the ab-plane.  
Therefore, we did not attempt to vary the c-components of the Tl and
O(3) positions in our model; they are fixed at the average positions given by
diffraction studies.  Furthermore, the  mean-square-root of the Tl
displacement is about 0.15 \Aa in the ab-plane ($\sqrt{U_{11}}$ or 
$\sqrt{U_{22}}$) and 0.09 \Aa along the c-axis ($\sqrt{U_{33}}$) 
in the refinement of diffraction
studies\cite{Torardi,Parise89,Shimakawa,Kolesnikov,Manako}, which are
quantitatively
consistent with our XAFS results.  Given the Tl displacement of 0.11 \Aa
in the ab-plane and the Tl Debye-Waller factor of 0.09 \Aa (see Table III),
the approximately
estimated mean-square-root of the Tl displacement in the ab-plane
is 0.14 \Aa  ($\sqrt{0.11^2+0.09^2}$ \AA), very colse to that of
the diffraction results. 

The model (Fig. 9) based on our XAFS data analyses suggests that the Tl
displacements are correlated over a short range.   The 
partial-pair-distribution-function (PPDF) for the Tl-Tl pairs
 extracted from our data
 shows a dip near the average distance as shown in Fig. 11a, which is
significantly different from the Tl-Tl PDF for the Tl in the ideal sites 
 as shown as the dotted curve in Fig. 11b (with a Debye-Waller factor 0.07 \AA).   
The second peak in the ideal structure is
split into two peaks in the extracted PDF, one near 3.7
\Aa and another one near 4.0 \AA, thus a dip is introduced near the average
distance 3.87 \AA.  When we fit the data using the ideal structural model
(4 Tl-Tl at 3.51 \AA, 4 Tl-Tl at 3.87 \Aa and 4 Tl-Ba at 3.89 \AA),  the
longer Tl-Tl pair is moved to 3.75 \Aa (solid curve in Fig. 11b) which
physically does not make sense because the average Tl-Tl distance in the
ab-plane should equal to the average lattice, 3.87 \AA.

Our model is similar to that suggested by the PDF analysis of the pulsed-neutron
scattering spectrum in Tl-2212\cite{Dmowksi,Toby89}, but more specific about the 
relative shift for the Tl atoms between
the two consecutive TlO layers.  Since the interlayer Tl-Tl distance
remains unchanged,  the displacement of the nearest Tl-Tl pairs between the TlO
layers should be either in the same direction or perpendicular to each
other, but not in opposite directions. The shifts of the
Tl and O(3) atoms and the pair distances obtained from XAFS, however,
are different from those obtained by PDF even if the small  changes in
the lattice constants between Tl-2201 and Tl-2212 are taken into account. 
The PDF analysis for Tl-2212 suggests that the Tl atom
shifts by 0.32 \Aa along the $<110>$ direction, which results in three split
Tl-Ba distances with $\delta r\simeq$0.23 \AA.  This is not consistent
with our Ba K-edge XAFS data ($\delta r\simeq$0.1 \AA).  We point out
that XAFS probes the local environment around each selected element
separately, and thus determines the average local structure around every different
element.  The combination of the information from different types of
absorbing atoms gives a more constrained result for the local structure.
On the other hand, PDF analysis uses all the atomic pairs in the material together,
so that identifying individual partial pair-distribution-functions (PPDFs) 
is a much more involved process.  Consequently, XAFS
offers an advantage in studying the near neighbor local structure around an
absorbing atom in complicated materials.

Some diffraction studies also revealed a Tl displacement from
its ideal site\cite{Parise89,Kolesnikov,Liu}.  An x-ray structural study of 
tetragonal single crystals
(Tl$_{1.85}$Cu$_{0.15}$)Ba$_2$CuO$_6$ with \Tc=110 K suggested that
 the Tl is displaced along the $<110>$ direction by 0.11 \AA\cite{Kolesnikov}.
Another x-ray diffraction study on tetragonal single crystals
Tl$_{1.87}$Ba$_2$Cu$_{1.13}$O$_{6+\delta}$ with \Tc=12 K showed a Tl 
displacement of 0.16 \Aa along the $<100>$ direction\cite{Liu}.  A neutron 
powder diffraction
investigation of orthorhombic Tl$_2$Ba$_2$CuO$_{6+\delta}$ with \Tc=90 K
inferred a Tl displacement of 0.14 \Aa along the $b$-axis\cite{Parise89}
 (which is along the $<110>$ direction in a tetragonal unit cell).  
In all these diffraction studies, the Tl atoms are assumed to be
statistically distributed in the displaced sites.  This leads to an uncorrelated
 Tl displacement with the Tl-Tl PDF having its  largest amplitude  
 at the average distance (see dotted lines in Fig. 11b and c), 
which is significantly different from the PPDF proposed by our XAFS study.  
  To further check
the  models suggested by diffraction studies, we have tried to fit 
the XAFS data with the uncorrelated
Tl displacement model for displacements
 along both the $<110>$ and $<100>$ directions.  

\subsubsection{ Tl displacement along the $<110>$ direction}

In the  $<110>$ direction Tl displacement  model suggested by one diffraction
study\cite{Kolesnikov}, the four-fold displaced sites of the Tl are
randomly occupied.  Thus, to check that model in the XAFS data analysis,  
eight Tl-Tl pairs must be included in the fit with distances $r_1-2\delta r_1$,
$r_1-\delta r_1$, $r_1$, $r_1+\delta r_1$, $r_1+2\delta r_1$, 
$r_2 -1.9\delta r_1$, $r_2$, $r_2+1.9\delta r_1$ and amplitudes
0.25, 1, 1.5, 1, 0.25, 1, 2, 1, respectively.  
Here, $r_1$ and $r_2$ are the average Tl-Tl distances between and within
the TlO layers, respectively; the splitting of the Tl-Tl distances from
their average distances are proportional to $\delta r_1$ to first order.
Correspondingly, three Tl-Ba
pairs must be included in the fit with distances $r_3+\delta r_3$, $r_3$,
$r_3-\delta r_3$ and amplitudes 1, 2, 1, respectively.  Three Tl-O pairs
near 2.0 \AA, 2.3 \Aa and 4.4 \Aa were also included.

 The fit was started with the Tl displacement of 0.11 \Aa from its ideal
site (which corresponds to $\delta r_1$=0.085 \Aa and $\delta r_3$=0.08
\AA) and Debye-Waller 
factors  of 0.07 \Aa for the Tl-Tl and Tl-Ba pairs.  
The starting Tl-Tl PDF is shown in Fig. 11b as dotted line.
To reduce the number of fitting parameters and to check the proposed
model, we fixed $r_1$, $r_2$, and $r_3$ at the average distances (3.51 \AA,
3.87 \AA,  and 3.89 \AA, respectively)
 given by diffraction study, and the amplitudes of the Tl and Ba peaks
to the numbers  predicted by the model.  The E$_o$ shifts for the same
type of back-scattering atoms  are kept equal except for the nearest O
atoms.  With these constraints, the total number of variables is 25 for
this 14-pair-fit.  The fitting range is exactly the same as that used for
the correlated model (Sec. V.C).

  The extracted PPDF is plotted in Fig. 11c as  the solid line.  The 
quality of this fit with 25 parameters is
comparable to the one obtained from the constrained fit using the correlated 
model (Fig. 9) with only 16 variables. 
The distances changed a little from the starting values to $\delta
r_1$=0.095 \Aa and $\delta r_3$=0.07 \AA. 
Since the weighted number of the
Tl-Tl pairs are fixed, their Debye-Waller factors are varied to adjust the
height of the peaks.  The extracted Debye-Waller factors
are less than 0.04 \Aa for the Tl-Tl peak at 3.51 \Aa and larger than 0.2 \Aa for  
the Tl-Tl peak at 3.87 \AA, which   results in a sharp peak at 
3.51 \Aa and a dip near 3.87 \Aa in the Tl-Tl PDF as shown in Fig. 11c by
the solid line.  For the Tl-Ba PDF, there is no difference between 
the $<110>$ direction correlated and uncorrelated  Tl displacements, and  
the extracted Tl-Ba PDFs are very similar.

Although the quality of the fit for the uncorrelated model (25 variables in
the fit) is comparable to that of the correlated model (16 variables in the fit),
  the extracted Tl-Tl PDF (solid line in Fig. 11c) from the constrained fit 
is significantly different from the starting Tl-Tl PDF ( dotted line in
Fig. 11c) which corresponds to the uncorrelated model.  Furthermore,
the extracted Tl-Tl PDF appears to be very close, within the imposed
constraints,  to  that
obtained from the correlated displacement model (solid line in Fig. 11a); 
{\it a dip shows up at the average distance 3.87 \AA.   For the
uncorrelated model, a peak must show up at the average distance in the Tl-Tl PDF}.
Clearly, our XAFS data do not support the uncorrelated $<110>$  Tl
displacement model.

\subsubsection{Tl displacement along the $<100>$ direction}

For the $<100>$ direction uncorrelated, four-fold Tl displacement model
suggested by diffraction study\cite{Liu},
eight Tl-Tl pairs must be included in the XAFS fit with distances 
$r_1-\delta r_1$,
$r_1$, $r_1+\delta r_1$, $r_2-1.77\delta r_1$, $r_2-0.885\delta
r_1$, $r_2$, $r_2+0.885\delta r_1$, $r_2+1.77\delta r_1$
and amplitudes 1.0, 2.0, 1.0, 0.25, 1.0, 1.5, 1.0, 0.25, respectively.
Two split Tl-Ba pairs are included in the fit with distances
 $r_3-\delta r_3$, $r_3+\delta r_3$.  Three Tl-O pairs are also included.

The fit was started with the Tl displacement obtained from diffraction,  
0.16 \AA\cite{Liu} (which corresponds to 
$\delta r_1=0.15 $ \Aa and $\delta r_3$=0.07 \AA).  As was done for the fit to 
the  $<110>$ uncorrelated  Tl displacement model, $r_1$, $r_2$, $r_3$, and 
the amplitudes of the Tl and Ba peaks are fixed.   The total number of 
variables is also 25.  The quality of the constrained fit is about 30\% 
worse than that of the fit for the
$<110>$ direction correlated displacement.
The Tl-Tl peak at 3.87 \Aa is
depressed by a large Debye-Waller factor, 0.2 \AA,  in the fitting process. 
The obtained distances, with $\delta r_1$=0.25 \Aa and $\delta r_3$=0.06 \AA, 
deviate significantly from the starting model and are no longer consistent
with the starting model. 
The extracted Tl-Tl PDF is plotted in Fig. 11d (solid
line), which clearly moves away from the starting Tl-Tl PDF suggested by
diffraction and towards the correlated Tl-Tl PDF (solid line in Fig.
11a).  

The extracted Tl-Ba PDF is somewhat similar to that obtained from the
$<110>$ direction displacement, but it is difficult
 to distinguish between
one broad peak, two split peaks, and three split peaks.

 The above analyses clearly show that
 the Tl-Tl peak at the average lattice distance, 3.87 \AA, is
essentially destroyed by the large Debye-Waller factor in the uncorrelated Tl
displacement model, for displacements along either the $<110>$ or the $<100>$ 
directions, and the extracted PPDF from the uncorrelated model tends to move 
away from 
the PPDF generated by the
starting model and  towards the PPDF obtained by the correlated model.
Therefore, we conclude
that our XAFS data do not support the uncorrelated Tl displacement along
either the $<110>$ direction or the $<100>$ direction proposed by the
diffraction studies\cite{Parise89,Kolesnikov,Liu}. 
We point out that XAFS is more sensitive
than diffraction to small local structural distortions which have no
long range order.  This can be seen from the following example.  Rietveld
refinements were performed for Tl-2212 using 7 distorted models with
different directions of displacement including two correlated
 PDF models\cite{Toby89}. 
The results of these refinements show relatively little sensitivity to
the choice of model used for the TlO layers.   Three models, {\it i.e.},
the uncorrelated four-fold displacement of both O(3) and Tl along the $<100>$ 
directions,  the uncorrelated four-fold displacement of O(3) along the $<100>$ 
and Tl along the $<110>$ directions, and the correlated two-fold displacement
of both O(3) and Tl along the $<110>$ directions 
(with the space group Pcca)\cite{Toby89}, result in essentially the
same quality of the fit to the data.  In the XAFS data analysis, these three 
models can easily be distinguished  as shown above.
  However, the model we proposed may not be unique.  
More complicated models may generate a similar PPDF to that shown in Fig. 11a.

\subsection{Comparisons of XAFS data for samples with different \Tc}
The general picture  presented in this paper for the distortions of the TlO 
layers is similar to that in Ref. 15, but with more extensive XAFS data 
for \tls and more  constraints on possible models.  For the  XAFS data
at the Tl edge, both the Tl-Tl and Tl-Ba peaks are split into several peaks 
which all overlap; at the Ba edge, only the Ba-Tl peak is split.  Therefore, 
the XAFS
data analysis at the Tl edge is more complicated and technically involved than 
at the  Ba edge.  Although the Ba edge XAFS data are not quite as sensitive to 
the Tl distortion as the Tl edge XAFS data, they help rule out some possible
models which fit the Tl edge XAFS data.

We also collected Tl L$_{III}-$edge XAFS data from another
sample\cite{Strom93}, Tl$_{1.85}$Ba$_2$CuO$_{y}$, with \Tc=76 K.  
This sample is
made in a similar  way as described in Sec. IIA.  It has a tetragonal
structure with lattice constants $a=3.8626$ \AA, $c=23.224$ \AA.  
We also have the Tl L$_{III}-$edge 
XAFS data used in the previous study\cite{LiTl}, which were
collected at 78 K from a tetragonal \tls sample with \Tc=89 K and
lattice constants $a=3.866$ \AA, $c=23.239$ \AA.
All the data were reduced in exactly the same way as described in Sec. III.

The Tl L$_{III}-$edge XAFS data are plotted in Fig. 12 for the three samples with 
different T$_cs$; solid lines, \Tc=60 K; dotted lines, \Tc=76 K; 
dashed lines, \Tc=89 K.  All the data can be fit with the model shown in
Fig. 9.  The difference in the extracted distances from the constrained fit,
 among three samples, is within 0.01 \Aa, although the amplitude of the
Tl-O peak near 2 \Aa varies significantly among the samples.   
This clearly shows that the 
local  Tl displacement in  the TlO layer is not sensitive to  the structural
symmetry  and the transition temperature \Tc.  This local distortion
is more likely intrinsic to the Tl-2201 compound, which will be further
discussed in the following section.

There is a significant difference in the amplitude of the first peak at
$r=2.04$ \Aa (c-axis Tl-O(2,3)).  The
amplitude was measured in two different ways.  First, the amplitudes were
obtained from the constrained fit using the FEFF5 standards, as described
earlier assuming a single peak for the first Tl-O pair.  
Second,  the data were fit one
to another in the range of  $1.0\sim 1.8$ \AA, and the relative amplitudes
 were obtained.  The average amplitude is plotted in Fig. 13 as a function
of \Tc;  it is very close to a linear function. The higher the Tl-O amplitude
of the peak near $r=2$ \AA, the lower \Tcs is.  This trend is
consistent with other experimental 
results\cite{Shimakawa,Shimakawa89,Kikuchi90};  a decrease in oxygen
content of about 0.1$\sim$0.15 per formula unit changed \Tcs from 0 to about
80$\sim$90 K\cite{Shimakawa89,Kikuchi90} although the absolute oxygen content 
was not well determined\cite{Shimakawa,Shimakawa89,Kikuchi90}.

The data plotted in Fig. 13 are from three samples with different Tl and O
contents which are not precisely determined.  It has been reported that 
both the Tl
and O contents can change \Tcs significantly\cite{Saini}.  The amplitude
of the Tl-O peak plotted in Fig. 13 is from the contributions of the
O(2) and the O(3) along the c-axis, and possibly an O(4)\cite{Shimakawa} 
if it exists.  
This amplitude can be affected in several ways: (a) a change in the O content 
at the O(2), O(3) and O(4)\cite{Shimakawa} sites, 
(b) the O  movement between the  O(3) and O(4) sites, (c)  changes in 
the Tl-O(2) and Tl-O(3) (along the c-axis) distances, (d)  a variation in 
the Tl content.
From the limited data we have, it is not clear what is the main reason
for the variation of the {\it measured} Tl-O  amplitude which seems much
too large 
to be explained solely by a  variation in O content.
However, Fig. 13 clearly shows that \Tcs is very sensitive to the O 
environment around the Tl atom.  It is generally believed that the \CuOs 
layer is the superconducting layer, thus this result also implies that a charge
transfer occurs between the \CuOs and TlO layers and it is mainly controlled
by the O arrangement around the Tl, not the Cu substitution for Tl.

\subsection{The Tl and Cu valences}

The atomic valence (V) can be estimated from the obtained local
structure using the Brown-Altermatt-Zachariasen
formulae\cite{Brown81,Brown85}, which describe the relationship
between the bond length ($r$) and the bond valence ($s$)
\begin{equation}
V=\Sigma s_i=\Sigma (r_i/r_o)^{-N}
\end {equation}
or
\begin{equation}
V=\Sigma s_i=\Sigma exp[(r_o-r_i)/B],
\end {equation}
where $r_o$, $N$ and $B$ are empirically determined
parameters\cite{Brown81,Brown85}.  
For Tl$^{3+}$-O, we used Eq. (2) with the most recent available parameters,
$r_o$=2.04 and $N$=6.4\cite{Otto93}  and the
structural model (Fig. 9) constructed based on our XAFS results for the
sample with the highest O concentration. The 
calculated Tl valence is (2.95$\pm$0.05)+, very close to the formal valence, 3+
(Tl$_{2}^{3+}$Ba$_{2}^{2+}$Cu$^{2+}$O$_{6}^{2-}$).
This is also consistent with the photoemission\cite{Nakajima89,Suzuki} and the
 x-ray-absorption near-edge spectra\cite{LiTl} measurements 
 which indicate a Tl valence slightly less than 3+.  If both the Tl 
and O(3) atoms were at their ideal sites ($4e$), the  calculated effective
valence for Tl would be 2.77+, which is too small compared to the formal
valence and not stable for the Tl ion (the Tl ion should be either Tl$^{1+}$
or Tl$^{3+}$).  This is probably part of the 
reason why the TlO layer is distorted. 

We also  calculated  the Cu valence,
 using Eq. (3) with $B=0.37$, $r_o=1.679$ \AA\cite{Brown85} and the
local structure derived from XAFS (Table I). The calculated Cu valence is
(2.15$\pm$0.05)+, which is
very close to the Cu(2) valence in \ybcs (2.20+)\cite{Brown89,Cava,Cava90} and
Cu in Tl$_2$Ba$_2$CaCu$_2$O$_8$ (2.11+)\cite{Hewat}
calculated using the same formula. 
Our estimated effective valences for Cu
and Tl are qualitatively in agreement with the charge transfer
(or local charge) picture between the TlO and \CuOs 
layers\cite{Cava},{\it i.e.}, Tl$_2$O$_3$ + CuO
$\longrightarrow$ [Tl$_2$O$_3$]$^{\delta -}$ + [CuO]$^{\delta +}$.
It is generally believed that the extra holes in the \CuOs are
predominantly at the O site\cite{Romberg,Fuggle,Emery}.  However, it
is not clear where the extra electrons are  in the TlO layers.  We note
that the extra holes in the \CuOs layer are not solely introduced by the 
reduction of the Tl valence.  The variation in O content can also change
the hole concentration in the \CuOs layer.  A decrease in O content of
about 0.1 per formula unit, which corresponds to a decrease in hole
concentration of 0.2, increases \Tcs from $\sim$0 to $\sim$90
K\cite{Shimakawa89,Kikuchi90}.  However, there is a controversy
in the literature about the absolute, optimum O content for samples with the
highest \Tcs \cite{Manako,Strom93,Shimakawa89,Kikuchi90}.

\subsection{The question of long range order}
Long range order modulations have been observed in Bi-Sr-Ca-Cu-O 
superconductors and related materials\cite{Tarascon91}.
Some long range order  in the TlO layers is also
observed in Tl-2201 and Tl-2212 by 
electron diffraction and  high-resolution transmission electron 
microscopy\cite{Parkin,Fitz,Oku,Beyers,Nakajima93}.  The modulation wave
vector is about $<-0.16,0.08,1>$ for tetragonal Tl-2201 ($a=b\simeq$ 3.8
\AA), $<0.08,0.24,1>$ for orthorhombic Tl-2201 ($a\simeq b \simeq$5.4 \AA),
and $<0.17,0,1>$ for tetragonal Tl-2212.  The superlattice for the first
case would have a large unite cell, $6a\times 12b$ in the ab-plane.
  However, these superlattices\cite {Parkin,Beyers}
are usually observed only with selected-area electron diffraction and are
much weaker in intensity than in the Bi-Sr-Ca-Cu-O superconductors
which exhibit strong supercell diffraction structure\cite{Tarascon91} and
c-axis displacemnets in the TlO layers are small (see Sec. VI.A).
Such superstructures are not always observed in the tetragonal Tl-2201
samples\cite{Nakajima93}.  Consequently, this long range order 
of the lattice distortions in the Tl-based superconductors seems weak, sample 
dependent, and may not be intrinsic.  Since both the Tl and O(3) atoms
can be shifted in two equivalent directions in the model we have proposed, 
material with this distorted local structure might  well have a 
superstructure over a limited range, which is qualitatively
consistent with the above electron diffraction measurements.  There are
many ways to construct  superstructures with periodicity of $6a\times 12b$
using the distorted $2a\times 2b$ unit cell shown in Fig. 9 for the TlO
layers.  Two possible examples that are probably over simplified
  are shown in Fig. 14.  Here the Tl 
displacements (not
shown) are the same from one $2a\times 2b$ cell to another, but the  
direction of the O displacements varies.  In  the upper panel,
the O positions are not periodic; some O atoms move together or apart,
while others move in the same direction.
In the lower panel, the direction of the O displacements is rotated by
90$^o$ in the right half of the supercell.

\section{Conclusions}

The local structure of \tls, with  an orthorhombic phase and \Tc=60 K,
 has been studied at the Cu K$-$, Ba K$-$, and
Tl L$_{III}-$edges using the XAFS technique.  Several new approaches have been
implemented, {\it
e.g.}, direct structural simulations up to 7 \Aa using the FEFF5 code, 
constrained fits, and  comparisons of PPDFs for different models.  

Both the qualitative and quantitative data analyses clearly show that
the Cu, O(1), O(2), and Ba atoms are at their ideal sites in the 
unit cell  as given by the diffraction studies, while the Tl and O(3) atoms
are displaced from the sites suggested by the average crystal structure.
The Tl-Tl distance at r=3.5 \Aa between the TlO layers remains, but the Tl-Tl
distance at 3.9 \Aa in the TlO layer is not observed and the Tl-Ba and Ba-Tl 
peaks are very broad.
The shorter Tl-O(3) distance in the TlO layer is about 2.33 \AA,
significantly
shorter than the distance calculated with both the Tl and O(3) atoms at
their ideal $4e$ sites ( $x=y=$0 or $\frac{1}{2}$).   An excellent fit
to the XAFS data can be achieved with a  correlated displacement model shown
in Fig. 9. 
The fitting results show that the Tl atom is displaced along the 
$<110>$ direction from
its ideal  site by about 0.11 \Aa and the O(3) atom is shifted  from
the $4e$ site by about 0.53 \Aa roughly along the $<100>$ direction.
This model also fits very well to the XAFS data collected from two tetragonal 
\tls samples with different T$_c$s.  

The XAFS data do not support the uncorrelated displacement models proposed 
by diffraction investigations\cite{Parise89,Kolesnikov,Liu}. Our model
is similar to that proposed by the PDF analyses of the neutron scattering  
data on Tl-2212\cite{Dmowksi}, but is more specific about the 
relative shift of the Tl atoms between the two consecutive TlO layers.
However, the Tl displacement, 0.11 \AA, obtained in our study, is only
one third of the value (0.32 \AA) suggested by the PDF. 

The estimated Cu and Tl valences imply a charge transfer between the \CuOs
and TlO layers which is consistent  with other experimental 
results\cite{LiTl,LiSr,Nakajima89,Suzuki}. The estimated Tl valence also
supports the distorted structure of the TlO layer obtained from our XAFS
analysis.  The estimated
Tl valence (2.95+) in the distorted TlO layer is much closer to its formal 
valence (3+) than that (2.77+) estimated from the ideal TlO layer. 

A comparison of the XAFS data at the Tl L$_{III}-$ edge from three \tls
samples clearly shows a correlation between the transition temperature and 
the local O environment around the Tl atoms.  This implies that the hole
concentration in the \CuOs layer is mainly controlled by the O arrangement
about the Tl in those samples, and not by  Cu substitution for the Tl.

\acknowledgments 

We thank  Corwin Booth for
help in the data collection.  The experiments were 
performed at the Stanford Synchrotron Radiation
Laboratory, which is operated by the U.S. Department of Energy, Division
of Chemical Sciences, and by the NIH, Biomedical Resource Technology Program,
Division of Research Resources.  The experiment is partially carried out on
UC/National Laboratories PRT beam time. 
The work is supported in part by NSF grant DMR-92-05204.
\widetext

\begin{figure}
\caption{The crystal structure for \tls showing the definitions for the
different O sites.  Half of the unit cell is shown
and the whole unit cell is symmetric about the \CuOs plane.  The dotted
lines indicate the atomic positions in the half unit cell.  The structural
parameters used here are from Ref. 4, in which the O(3) atom is displaced
from the $4e$ site along the a-axis by 0.37 \AA. \label{fig1}}
\end{figure}

\begin{figure}
\caption{The k-space XAFS data at the Cu K-, Ba K-, and Tl L$_{III}$-edges
 in \tl.\label{fig2}}
\end{figure}

\begin{figure}
\caption{(a-b) The absorptance above the Ba K-edge (a) and the Tl
L$_{III}$-edge (b) in \tl; solid lines, experimental data; dotted lines,
fits to the data using cubic splines; dashed lines,  fits to the data using
cubic splines plus the extracted background from BaO and Tl$_2$O$_3$,
respectively.  The dotted lines and
dashed lines are used to simulate the ``free-atom absorptance". 
(c-d) The Fourier transform of the Ba K-edge (c) and the Tl
L$_{III}$-edge (d) XAFS data from \tl, using the ``free-atom absorptance"
obtained from cubic spline fits (dotted line) and cubic spline plus
corresponding extracted background fits (solid lines), respectively. 
The transform ranges are over 3.5 to 15.5 \Ia at the Ba edge and 
3.3 to 15.5 \Ia at the Tl edge, and both are Gaussian
broadened by 0.3 \IA. The fast oscillatory curves are the real parts of the
complex transforms and the envelope curves are the magnitudes of the
transforms. \label{fig3}}
\end{figure}

\begin{figure}
\caption{The Fourier transform of the Cu K-edge XAFS, $k\chi (k)$, to real
space for \tl. The transform range is over 3.1 to 16.5 \IA, Gaussian
broadened by 0.3 \IA. The fast oscillatory curve is the real part of the
complex transform and the envelope curve is the magnitude of the transform.
The solid lines are the experimental data and the dotted lines are the
{\it simulations} (without a fit)
 using the FEFF5 code with the average crystallographic
structure\protect\cite{Torardi} including all the pairs within 7.5 \AA. \label{fig4}}
\end{figure}

\begin{figure}
\caption{The Fourier transform of the Ba K-edge XAFS, $k\chi (k)$, to real
space for \tl. The transform range is over 3.5 to 15.5 \IA, Gaussian
broadened by 0.3 \IA. The fast oscillatory curve is the real part of the
complex transform and the envelope curve is the magnitude of the transform.
The solid lines are the experimental data and the dotted lines are the
{\it simulations} using the FEFF5 code, (a) with the average crystallographic
structure\protect\cite{Torardi}, and (b) with the local distorted structural
model (Fig. 9), including all the pairs within 7.5 \AA. 
Part (c) shows a small section of the $x-$axis
enlarged:  experimental data (solid lines), 
simulations from part (a) (dotted lines), and simulations from
part (b) (dots). \label{fig5}}
\end{figure}

\begin{figure}
\caption{The Fourier transform of the Tl L$_{III}$-edge XAFS, $k\chi (k)$, 
to real space for \tl. The transform range is over 3.3 to 15.5 \IA, Gaussian
broadened by 0.3 \IA. The fast oscillatory curve is the real part of the
complex transform and the envelope curve is the magnitude of the transform.
The solid lines are the experimental data and the dotted lines are the
{\it simulations} using the FEFF5 code, (a) with the average crystallographic
structure\protect\cite{Torardi}, and (b) with the local distorted structural
model (Fig. 9), including all the pairs within 7.5 \AA.
Note that no fit of the simulated XAFS to the experimental data has been
made (for more information, see text).  \label{fig6}}
\end{figure}

\begin{figure}
\caption{The Fourier transform of $k\chi (k)$ at the Cu K-edge for \tl.
The transform range is the same as in Fig. 4.  Both the real part  
and  the magnitude of the transform are plotted.
The solid lines are the experimental data and the dotted lines are the
fits to the data.  The fitting range is 1.0 to 3.9 \AA.\label{fig7}}
\end{figure}

\begin{figure}
\caption{The Fourier transform of $k\chi (k)$ at the Ba K-edge for \tl.
The transform range is the same as in Fig. 5.  Both the real part  
and  the magnitude of the transform are plotted.
(a) The comparison between the experimental curves and fits to the
data. 
 The fitting range is 1.2 to 3.8 \AA.
The solid lines are the experimental data and the dotted lines are the
fits. 
(b) The contributions of the Ba-O, -Cu, and -Ba peaks in the fit.
(c) The contributions of the Ba-Tl pairs.  The top trace is the sum of the
Ba-Tl peaks, followed by the three individual peaks.\label{fig8}}
\end{figure}

\begin{figure}
\caption{The projection of the local structural model in the ab-plane for the
distorted TlO layers. Two successive layers are shown.  The best overall 
agreement between the model and the XAFS results
is achieved when the Tl atom is shifted along the $<110>$
direction by 0.11 \Aa and the O atom in the TlO layer is shifted along the
$[100]$ and/or $[010]$ directions by about 0.53 \Aa from its ideal site.
The O displacement along the $[010]$ direction is not shown in the figure.  
Note that the displacements of the Tl and O atoms are {\it correlated}, not
random.
\label{fig9}}
\end{figure}

\begin{figure}
\caption{The Fourier transform of $k\chi (k)$ at the Tl L$_{III}$-edge for \tl.
The transform range is the same as in Fig. 6.  Both the real part  
and  the magnitude of the transform are plotted.
(a) The comparison between the experiment results and fits to the data.
 The fitting range is 1.2 to 4.2 \AA.
The solid lines are the experimental data and the dotted lines are the
fits to the data. 
(b) The contributions of the Tl-O peaks in the fit.
(c) The contributions of the Tl-Tl pairs.  The top trace is the sum of the
Tl-Tl peaks, followed by the three individual peaks.
(d) The contributions of the Tl-Ba pairs.  The top trace is the sum of the
Tl-Ba peaks, followed by the three individual peaks.\label{fig10}}
\end{figure}

\begin{figure}
\caption{The Tl-Tl pair-distribution-function (PDF) from
(a)  the correlated Tl displacement model (refer to Fig. 9),  (b)
  the ideal structure, (c) the uncorrelated Tl displacements along
the $<110>$ direction,  and (d) the uncorrelated Tl displacements along
the $<100>$ direction. The dotted lines are the starting PDFs 
 and solids are the extracted  Tl-Tl PDFs from XAFS using  the corresponding 
structural models and  the constrained fits.
For more information, see Sec. VI.A.  \label{fig11}}
\end{figure}

\begin{figure}
\caption{The Fourier transform of the Tl L$_{III}$-edge XAFS, $k\chi (k)$, 
to real space. The transform range is over 3.2 to 15.5 \IA, Gaussian
broadened by 0.3 \IA. Both the real part and the magnitude of the Fourier 
transform are plotted.  The data are collected at T=80$\pm$2 K from three 
samples with \Tc=60 K (orthorhombic, solid lines), \Tc=76 K (tetragonal, dotted 
lines), and \Tc=89 K (tetragonal, dashed lines), respectively.
  For more information, refer to Sec. VI.B.  \label{fig12}}
\end{figure}

\begin{figure}
\caption{The  amplitude of the nearest Tl-O peak near $r=2$ \Aa  
versus \Tcs for three samples presented in Fig. 12. \label{fig13}}
\end{figure}

\begin{figure}
\caption{ Two possible models for the $6a\times 12b$ superlattice.  The Tl
displacements are the same from one $2a\times 2b$ unit cell to another while
the O displacements vary. \label{fig14}}

\end{figure}
\eject
\begin{table}
\setdec 00.000
\caption{Cu local structure obtained from XAFS and  
diffraction. 
``nbrs" 
indicates the weighted 
number of neighbors. The estimated errors are:
number of neighbors, $ \pm $ 15 \%; distances, $ \pm $ 0.02 \AA.}
\begin{tabular}{ccccccc}
Pairs&\multicolumn{3}{c}{XAFS}& &\multicolumn{2}{c}{Diffraction}\\
     & r(\AA) & nbrs&$\sigma$(\AA)&   &  r(\AA) & nbrs \\
\tableline
Cu-O(1)& 1.93 & 3.8  & 0.038 &  & 1.93 & 4 \\
Cu-O(2)& 2.71 & 1.9  & 0.093 &  & 2.72 & 2 \\
Cu-Ba  & 3.35 & 7.7  & 0.053 &  & 3.35 & 8 \\
Cu-Cu  & 3.86 & 3.2  & 0.046 &  & 3.87 & 4 \\
Cu-O(1)& 4.31 & 7.7  & 0.056 &  & 4.32 & 8 \\
\end{tabular}
\label{table1}
\end{table}

\begin{table}
\setdec 00.000
\caption{Ba local structure from XAFS, 
diffraction,
and the model.  The values listed in the brackets are from other diffraction
studies\protect\cite{Torardi,Parise88,Hewat88,Parise89,Shimakawa,Kolesnikov,Liu,Manako}.
 ``nbrs" 
indicates the weighted number of neighbors.
  In the
model, the Tl atoms are shifted from the $4e$ site along the $<110>$ direction by 
0.11 \Aa and the O(3) atoms are shifted from the $4e$ site by 0.53 \Aa 
 along the $<100>$ direction.  The estimated errors are:
number of neighbors, $ \pm $ 15 \%; distances, $ \pm $ 0.02 \AA.}
\begin{tabular}{cccccccccc}
Pairs&\multicolumn{3}{c}{XAFS}& &\multicolumn{2}{c}{Diffraction}&&\multicolumn{2}{c}{Model}\\
     & r(\AA) & nbrs&$\sigma$(\AA)&   &  r(\AA) & nbrs &  & r(\AA) & nbrs\\
\tableline
Ba-O(1)& 2.73 & 4.0 & 0.046 &  & 2.74 & 4 & &     &    \\
Ba-O(2)& 2.83 & 4.0 & 0.054 &  & 2.84 & 4 & &     &    \\
Ba-O(3)& 2.97 & 1.0 & 0.040 &  &2.95 (2.95-3.00)&1&& 2.98 & 1  \\
Ba-Cu  & 3.35 & 4.1 & 0.053 &  & 3.35 & 4 & &     &    \\
Ba-Ba  & 3.87 & 6.0 & 0.066 &  & 3.87 & 5 & &     &    \\
Ba-Tl  & 3.80 & 1.0 & 0.060& & 3.89 & 4 & &3.81 & 1  \\
       & 3.90 & 2.0 & 0.060& &      &   & & 3.89& 2   \\
       & 4.00 & 1.0 & 0.060& &      &   & &3.97 & 1 \\
\end{tabular}
\label{table2}
\end{table}

\begin{table}
\setdec 00.000
\caption{Tl local structure obtained from XAFS, 
diffraction,
and the model.  The values listed in the brackets are from other diffraction
studies\protect\cite{Torardi,Parise88,Hewat88,Parise89,Shimakawa,Kolesnikov,Liu,Manako}.
  ``nbrs" 
indicates the weighted number of neighbors.  In
the
model, the Tl atoms are shifted from the $4e$ site along the $<110>$ direction by 
0.11 \Aa and the O(3) atoms are shifted from the $4e$ site by 0.53 \Aa 
 along the $<100>$ direction.  The estimated errors are:
number of neighbors,  $ \pm $ 15 \%; distances, $ \pm $ 0.02 \AA.}
\begin{tabular}{cccccccccc}
Pairs&\multicolumn{3}{c}{XAFS}& &\multicolumn{2}{c}{Diffraction}&&\multicolumn{2}{c}{Model}\\
     & r(\AA) & nbrs&$\sigma$(\AA)&   &  r(\AA) & nbrs &  & r(\AA) & nbrs\\
\tableline
Tl-O(2,3)& 2.04 & 1.8 & 0.045&  & 1.99 (1.98-2.01)& 1 & &2.00& 1 \\
	 &      &     &      &  & 2.05 (2.01-2.09)& 1 & &2.09& 1 \\
Tl-O(3)  & 2.33 &(2.6)& 0.131&  & 2.43-3.07 (2.31-3.22) & 4  & &2.29-2.38& 2 \\
	 &      &     &      &  &                       &    & &3.12-3.24& 2\\
Tl-O(2)  & 4.39 & 5.4 & 0.117  &  & 4.35 & 6  &    &4.28-4.44& 6 \\
Tl-Tl    & 3.51 & 4.2 & 0.081&  & 3.51 & 4  & &3.51&2\\
	 &      &     &      &  &      &    & &3.52&2\\
	 & 3.72 & 2.1 & 0.069&  & 3.87 & 4  & &3.72&2\\
	 & 4.03 & 2.1 & 0.114&  &      &    & &4.02&2\\
Tl-Ba    & 3.82 & 0.95 & 0.060&  & 3.89 & 4  & &3.81&1\\
	 & 3.90 & 1.9 & 0.060&  &      &    & &3.89&2\\
	 & 3.98 & 0.95 & 0.060&  &      &    & &3.97&1\\
\end{tabular}
\label{table3}
\end{table}

\epsfxsize=6.5 in
\epsffile{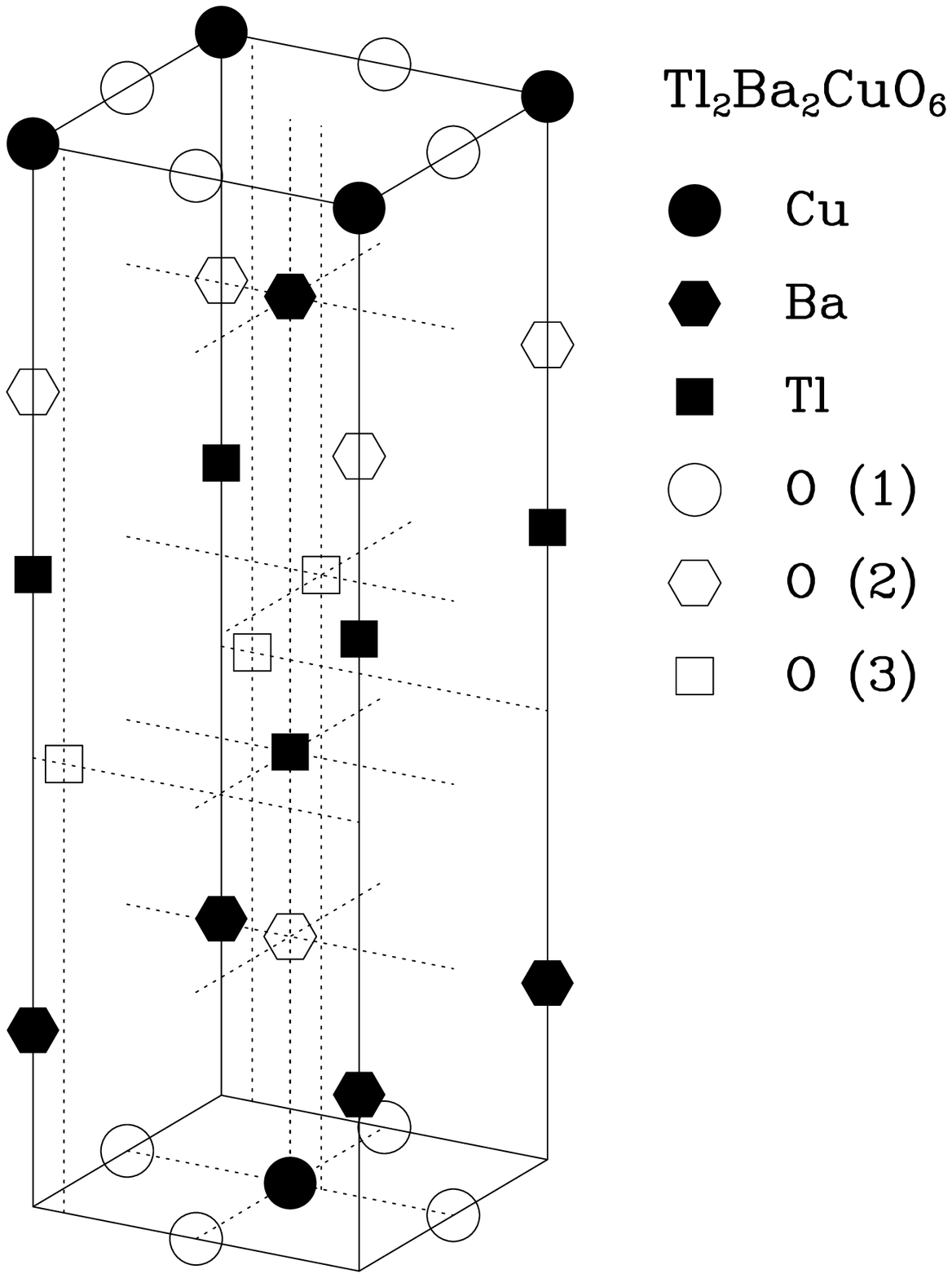}
\epsfxsize=6.5 in
\epsffile{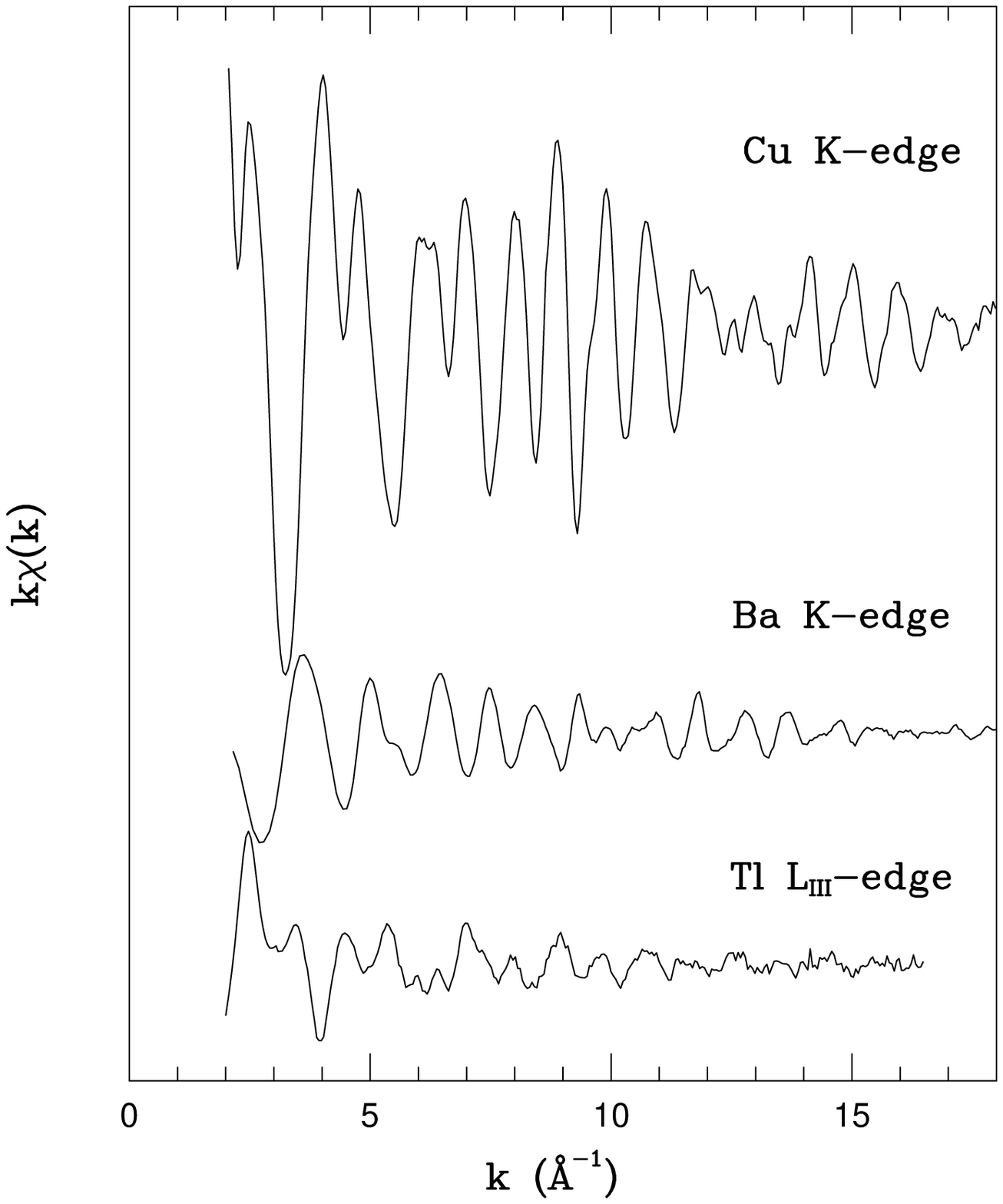}
\epsfxsize=6.5 in
\epsffile{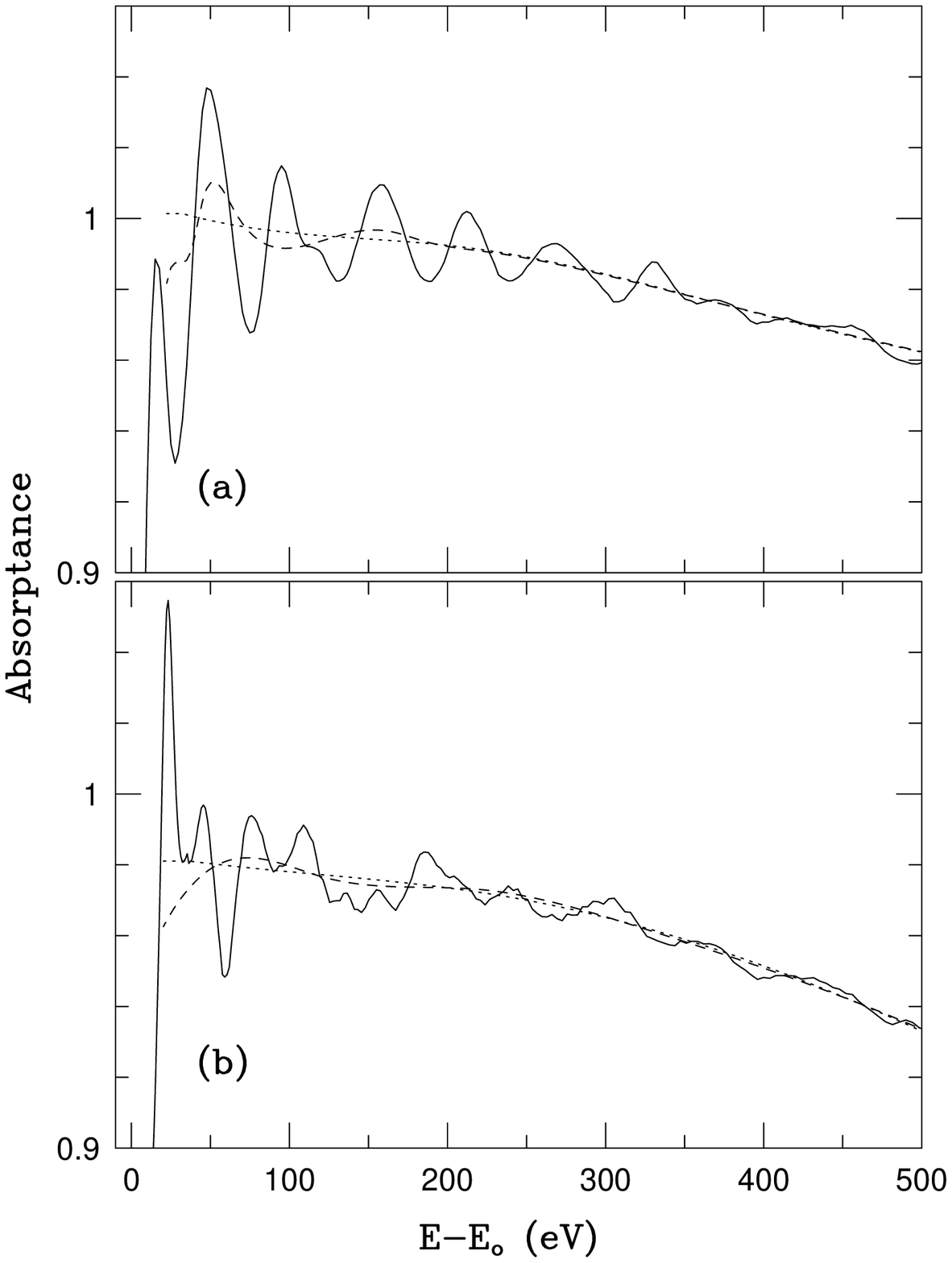}
\epsfxsize=6.5 in
\epsffile{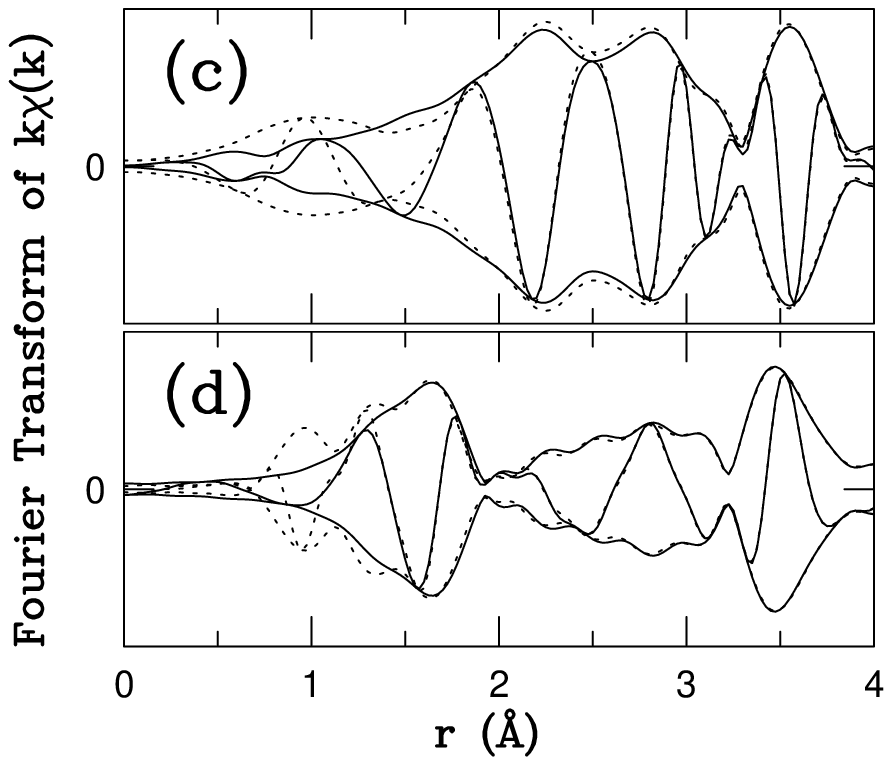}
\epsfxsize=6.5 in
\epsffile{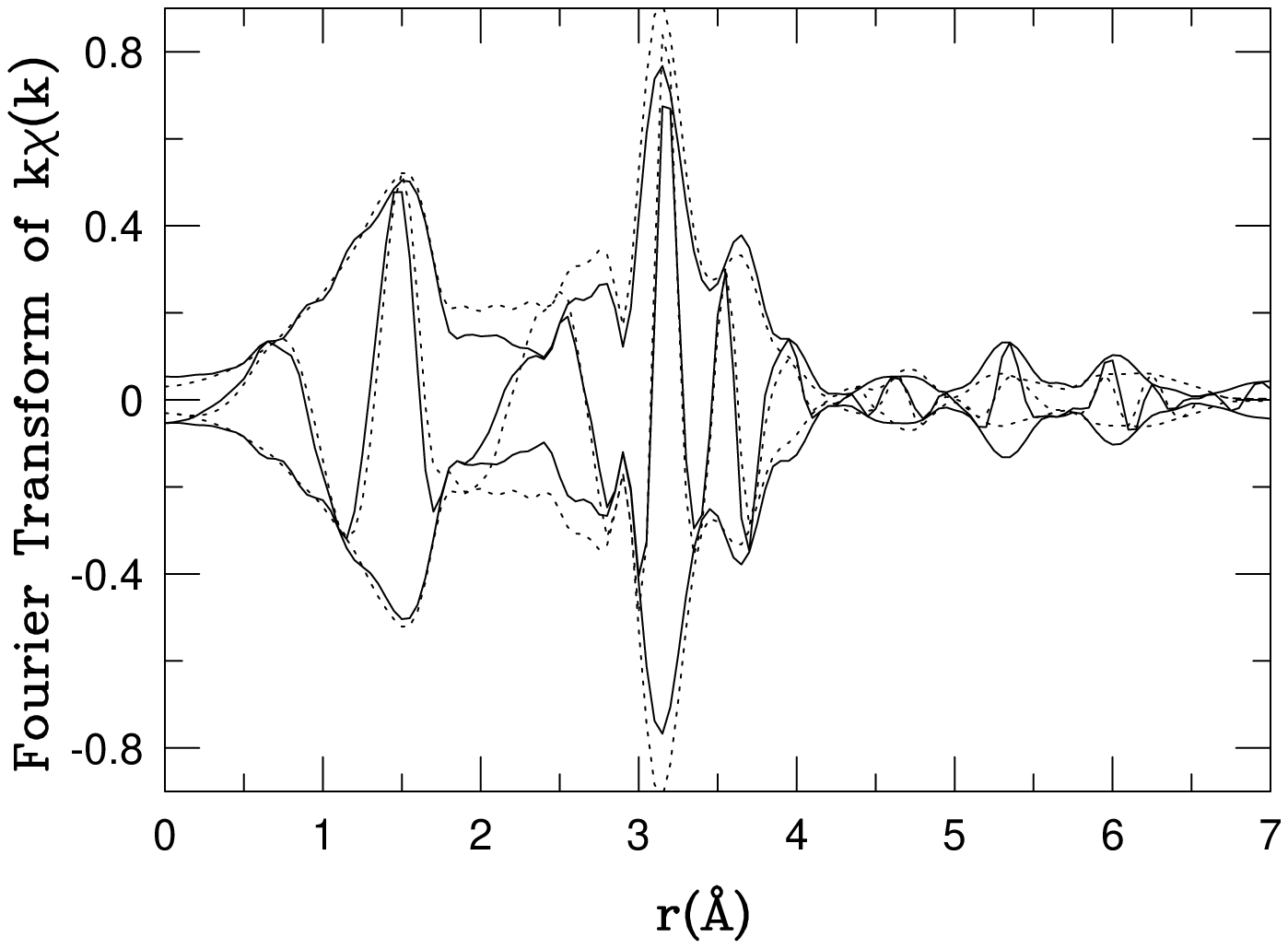}
\epsfxsize=6.5 in
\epsffile{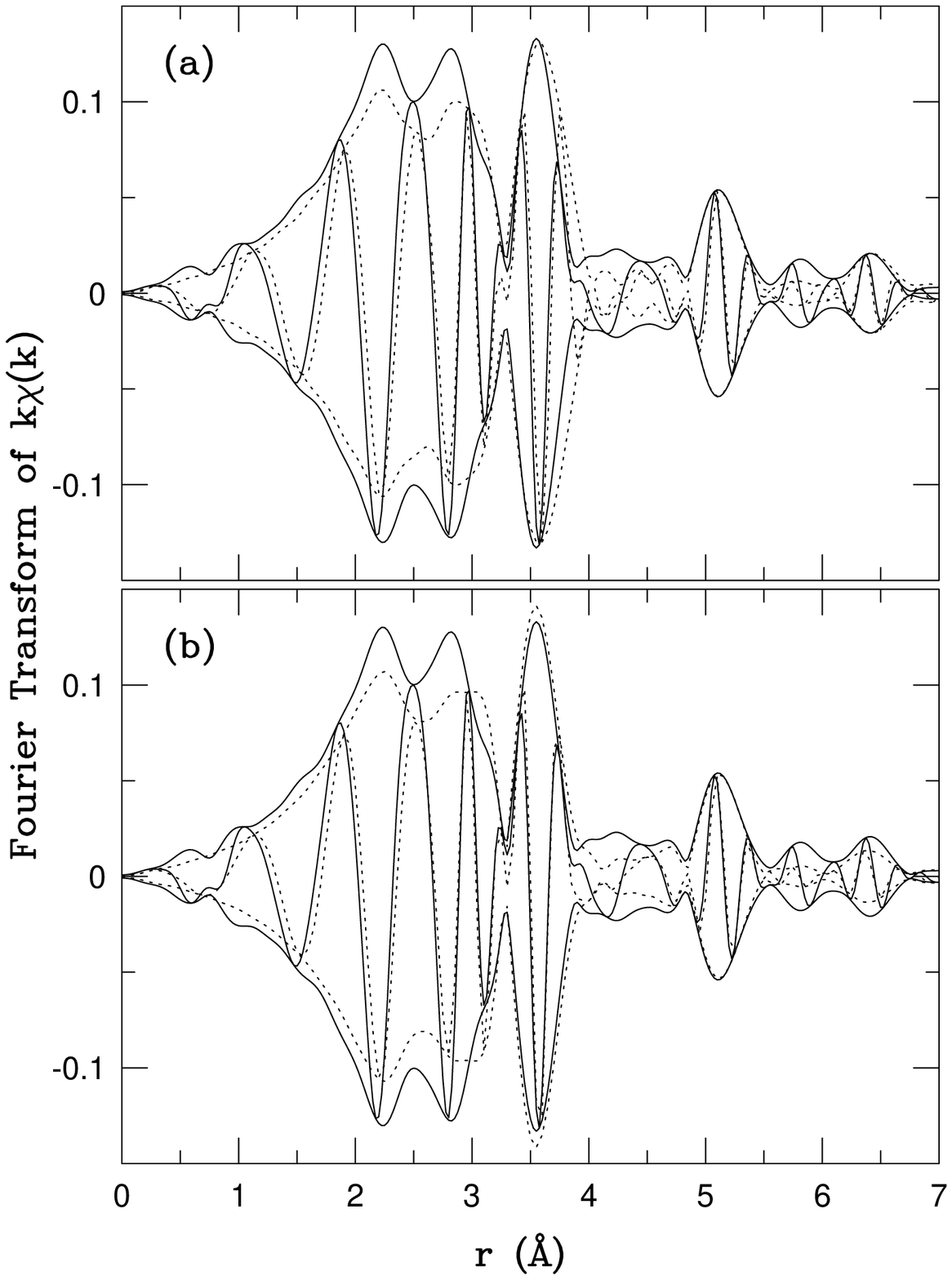}
\epsfxsize=6.5 in
\epsffile{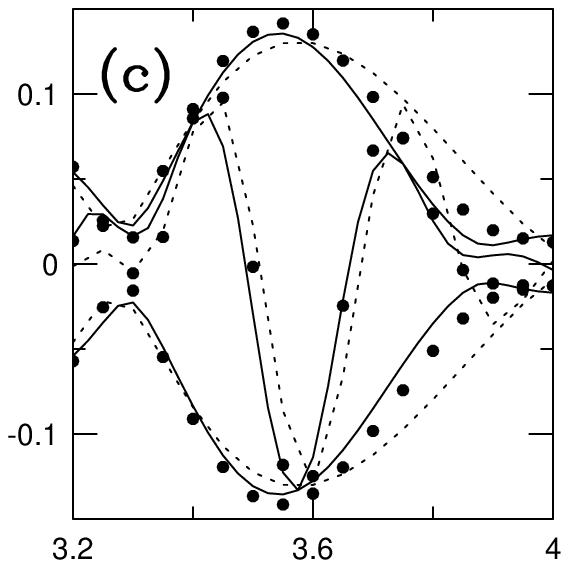}
\epsfxsize=6.5 in
\epsffile{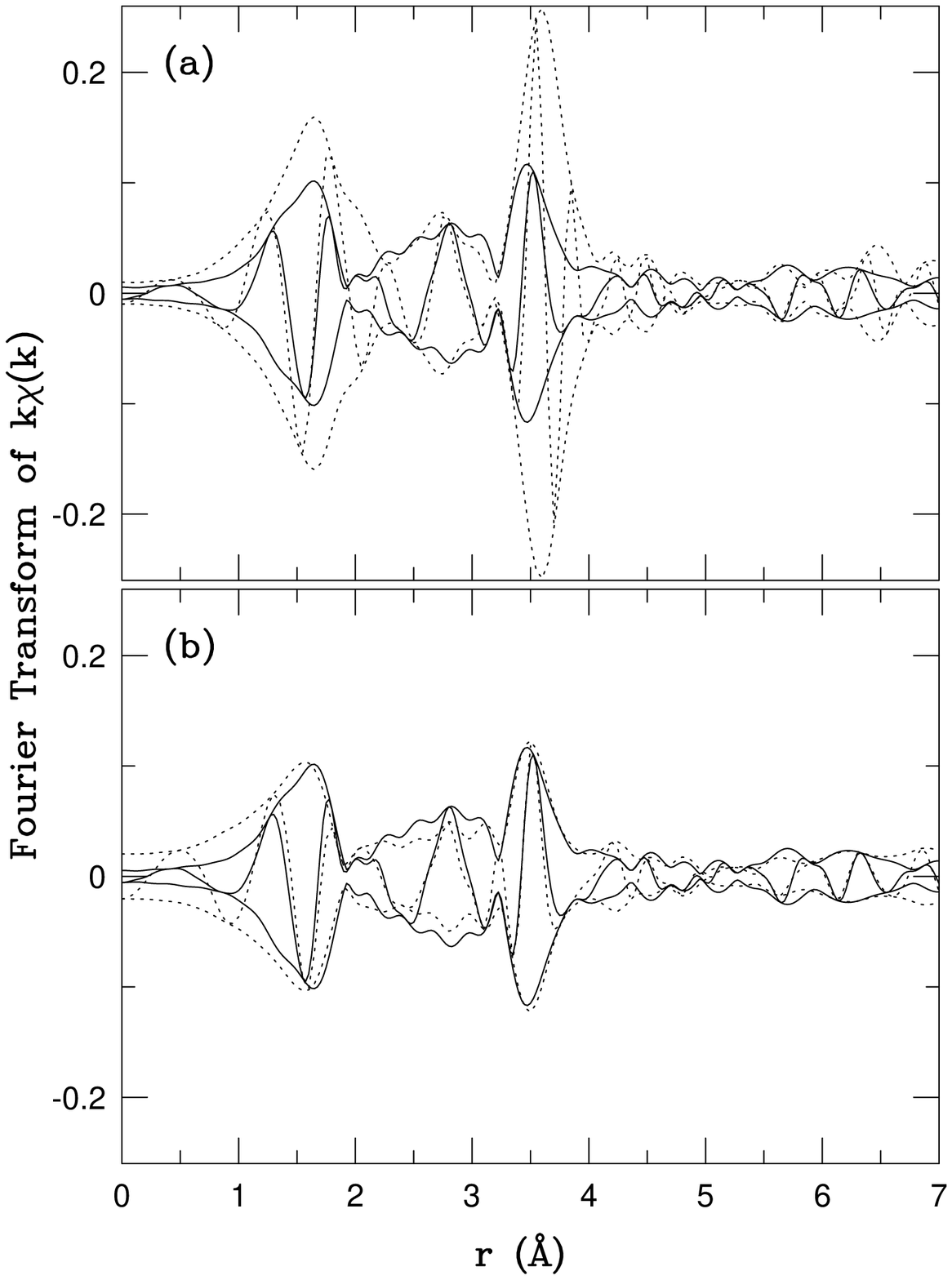}
\epsfxsize=6.5 in
\epsffile{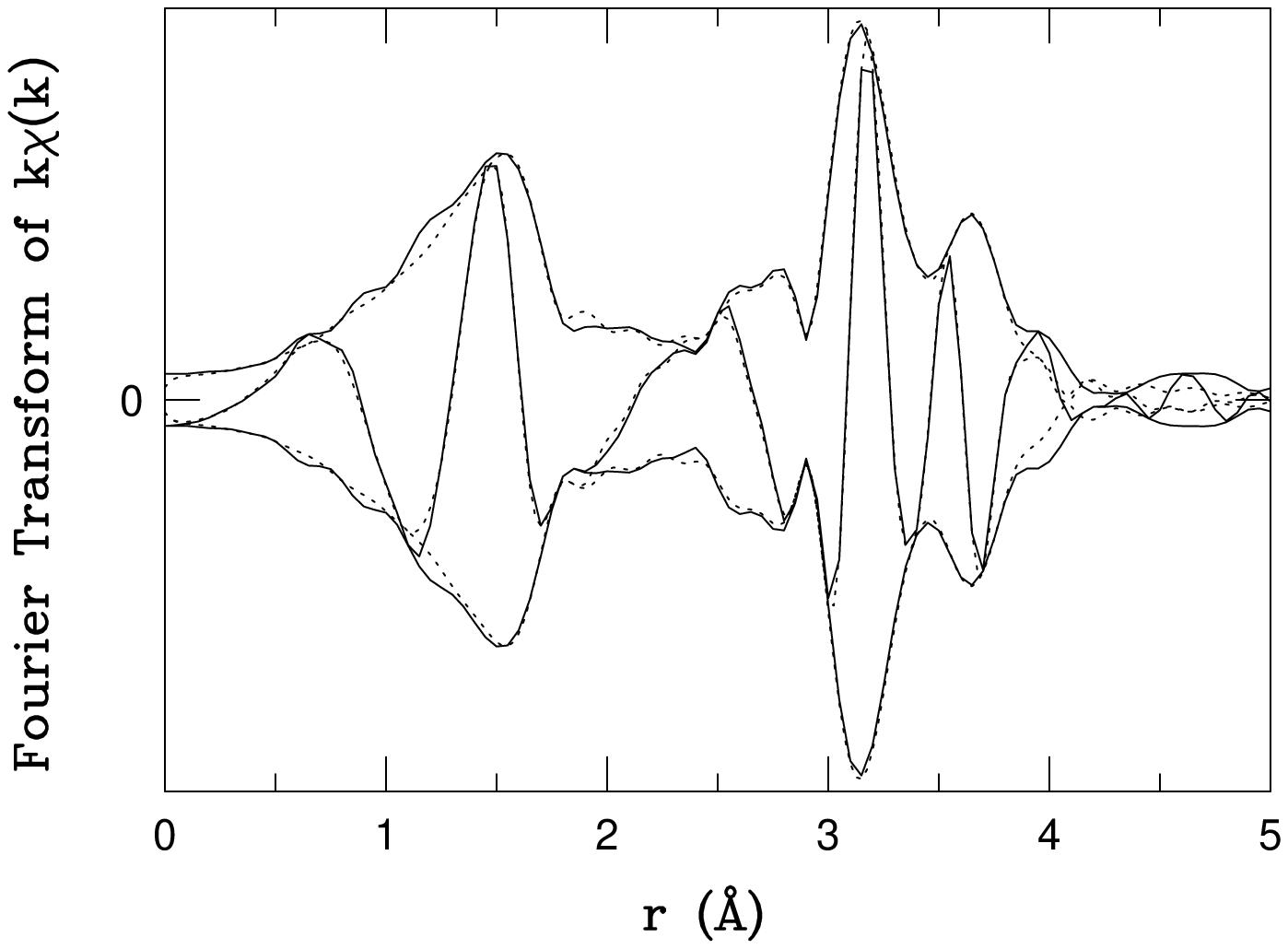}
\epsfxsize=6.5 in
\epsffile{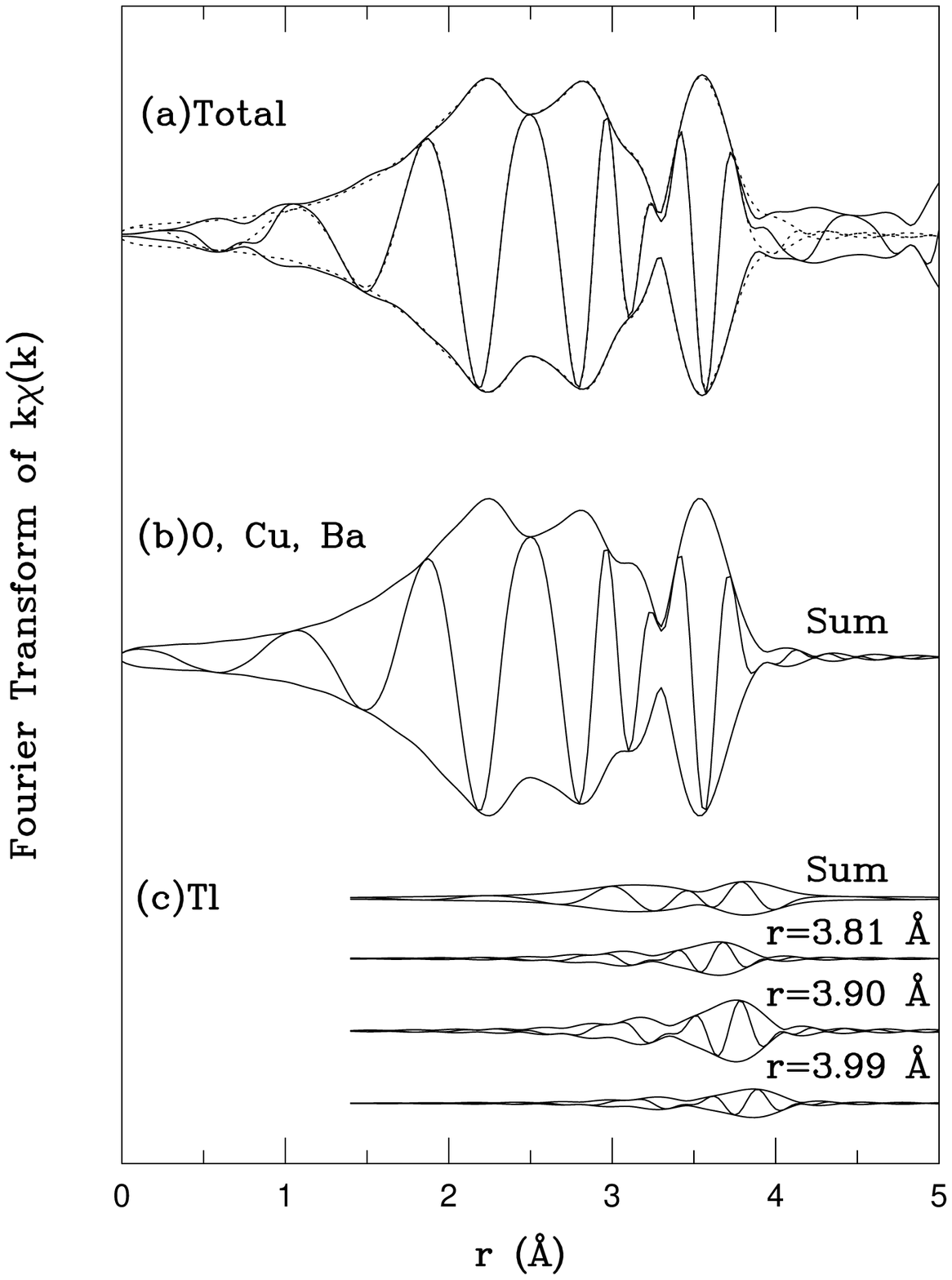}
\epsfxsize=4.5 in
\epsffile{fig.9.xfig.ps}
\epsfxsize=6.5 in
\epsffile{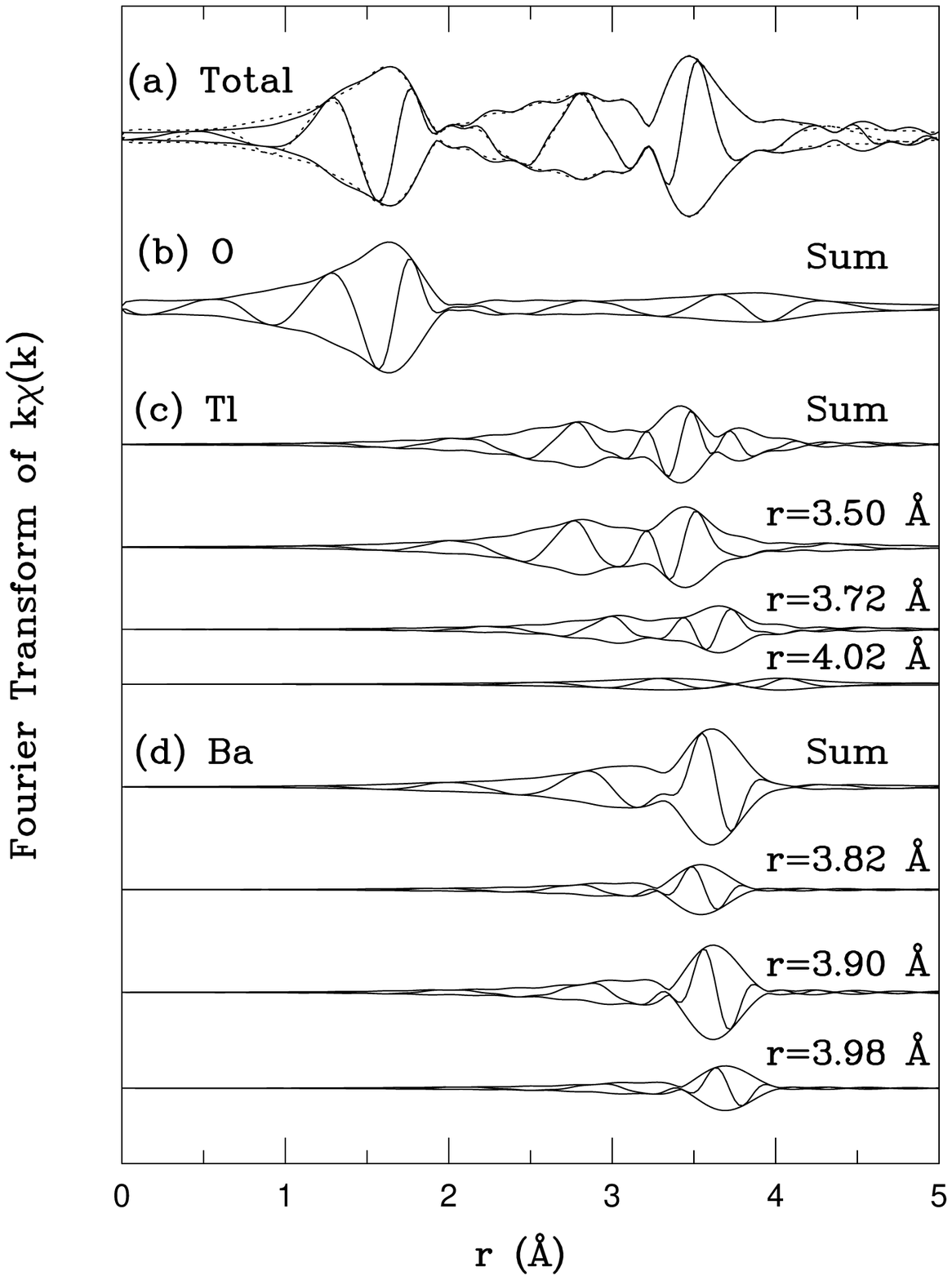}
\epsfxsize=6.5 in
\epsffile{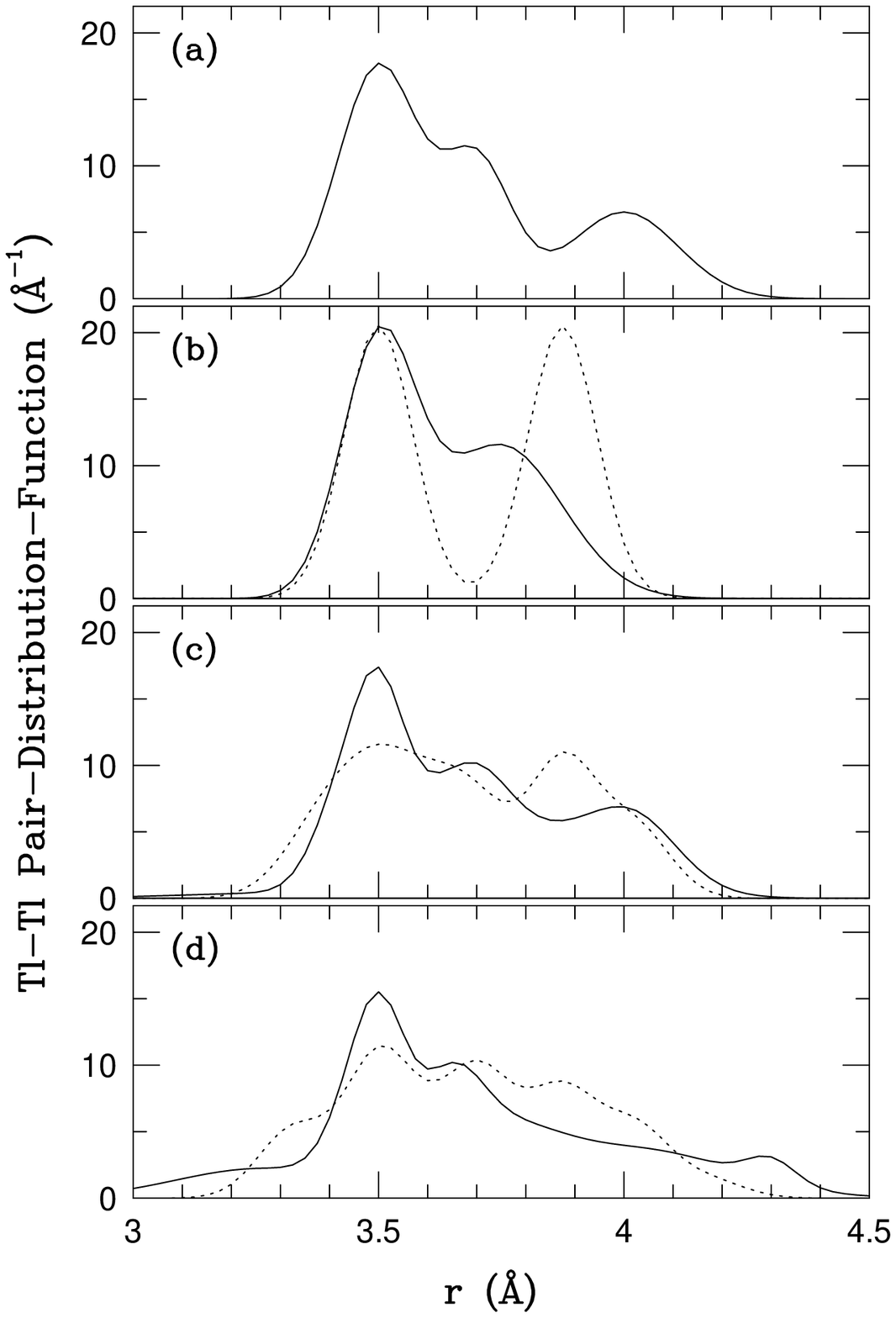}
\epsfxsize=6.5 in
\epsffile{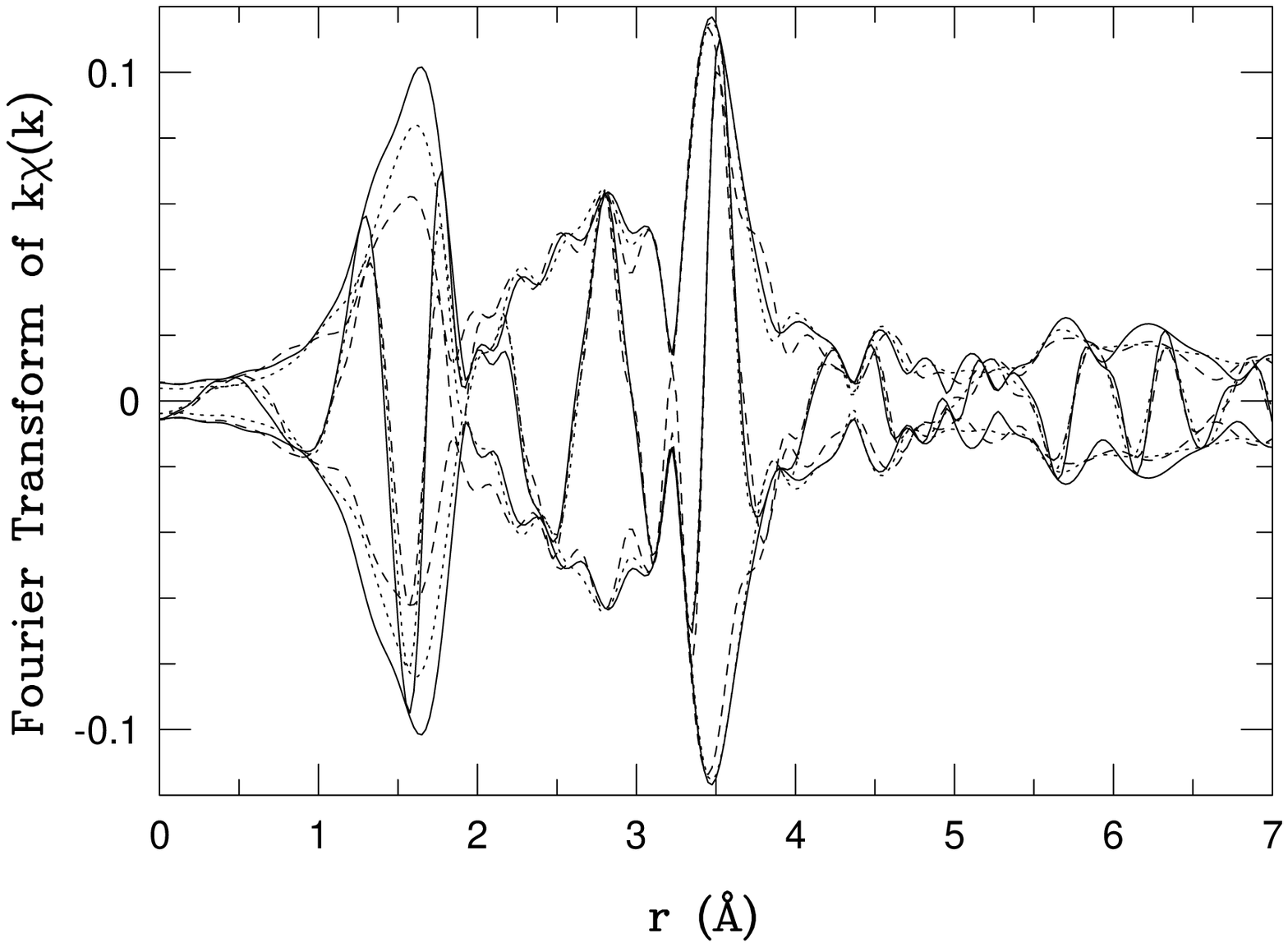}
\epsfxsize=6.5 in
\epsffile{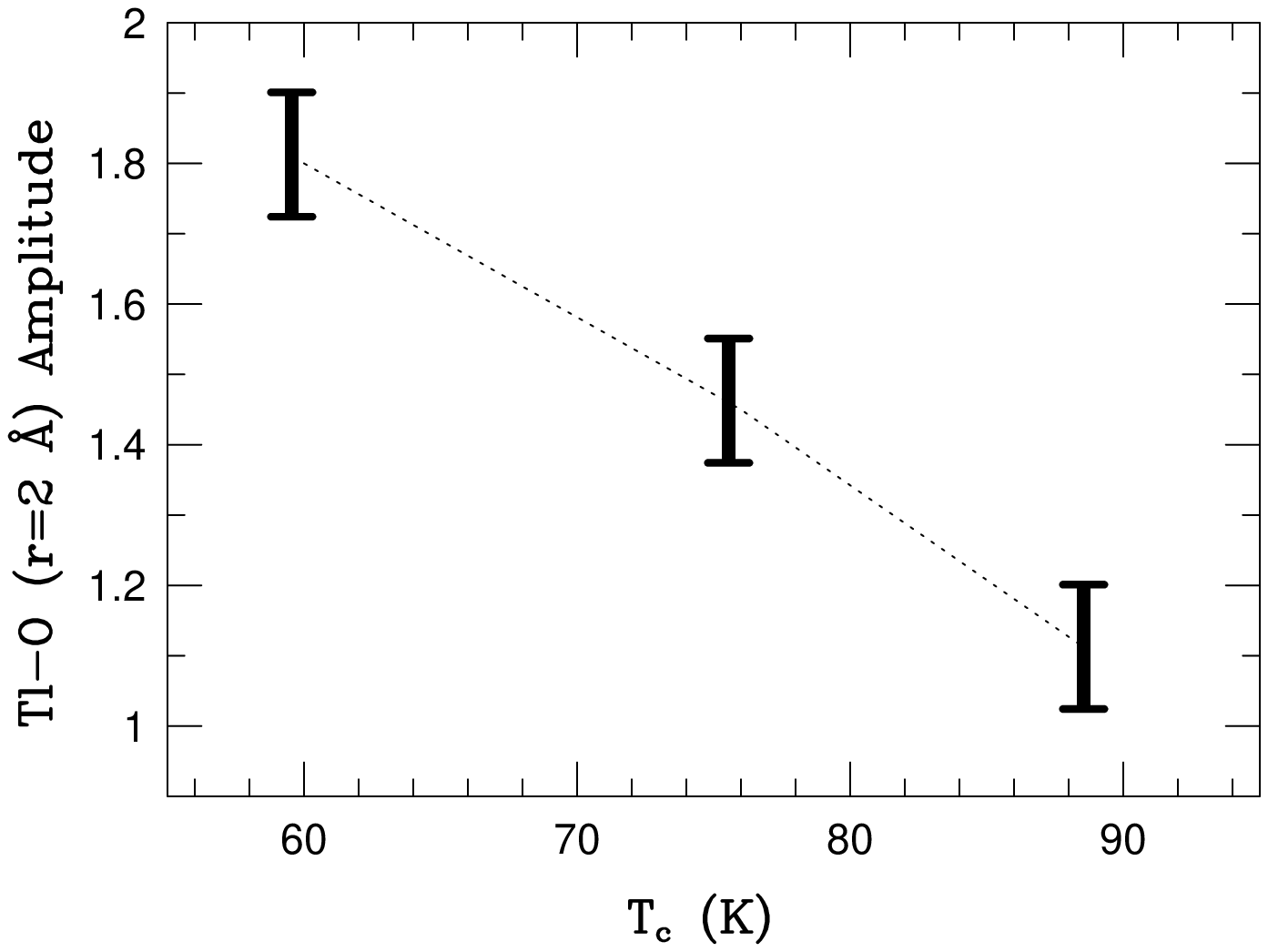}
\epsfxsize=6.5 in
\epsffile{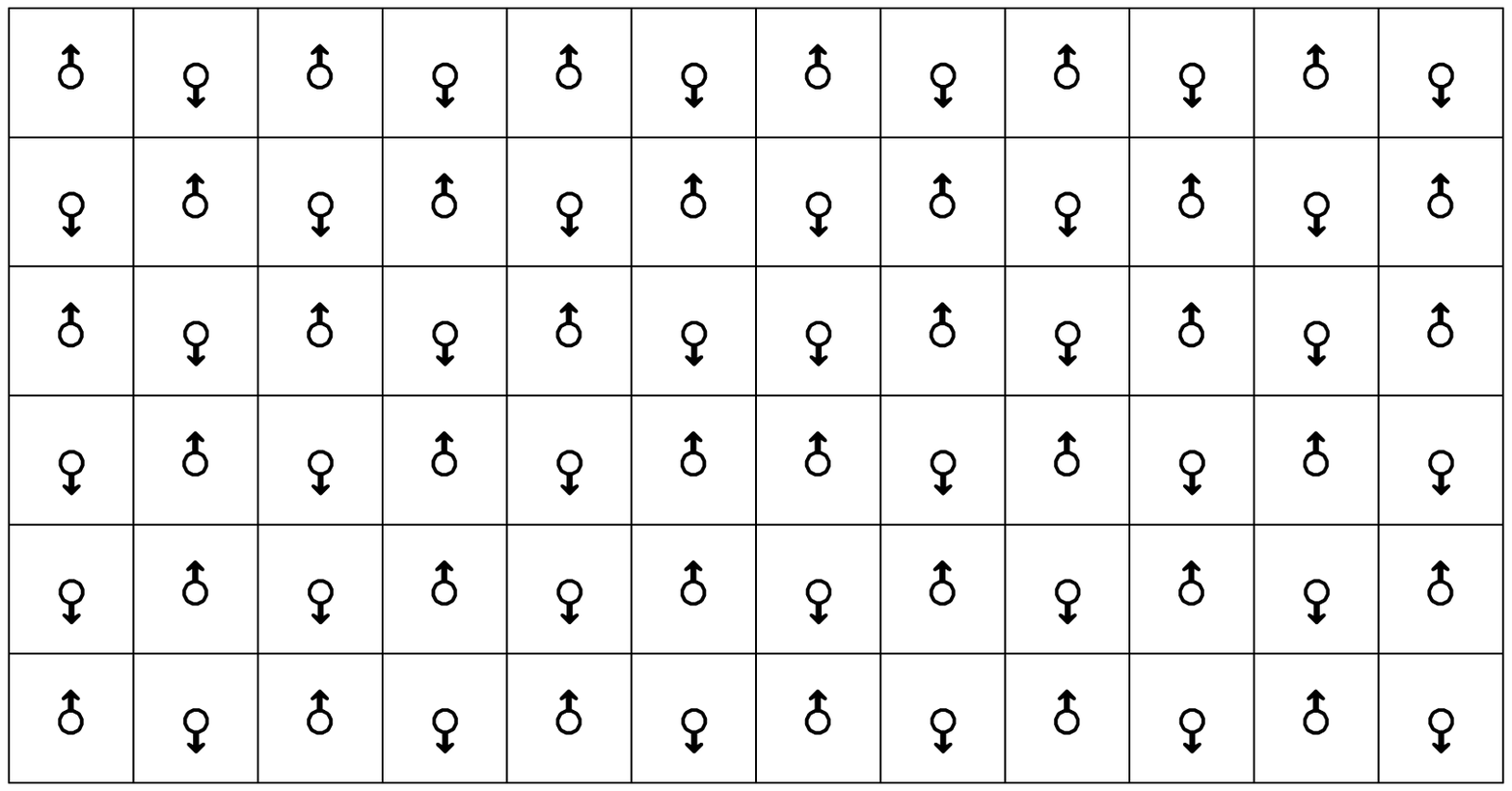}
\epsfxsize=6.5 in
\epsffile{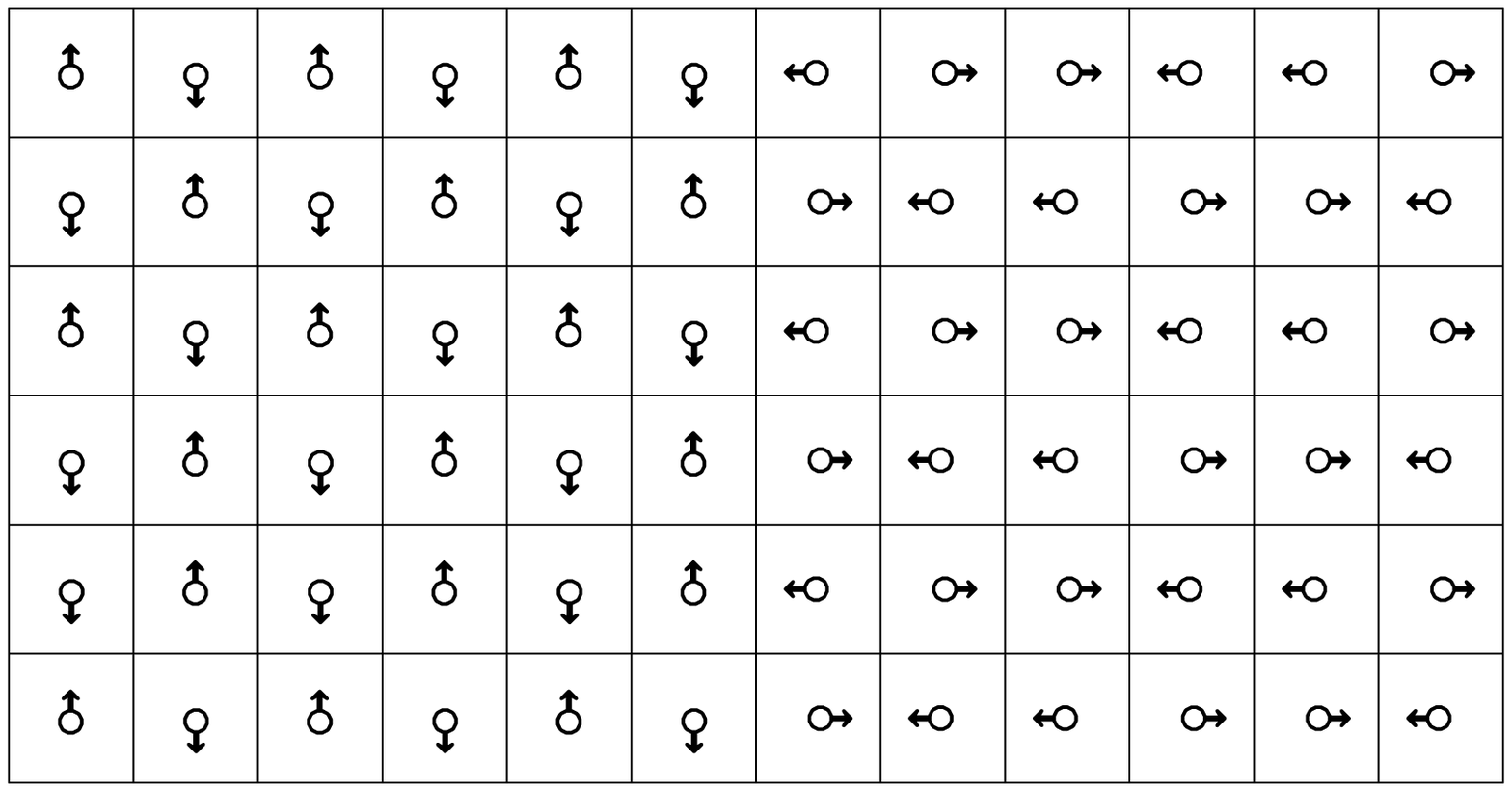}


\begin{references}
\bibitem{Note} In this paper, we have used $S_o^2$=0.9 for the Tl-Tl pair.
In Ref. 9, $S_o^2$ was set to be about 0.5 which seems too small according to
our recent detailed comparison of FEFF5 calculation with experimental
XAFS\cite{LiFeff}.
\end{references}
\end{document}